\documentclass[preprint2]{aastex}

\newcommand{\VVI}{$V$--($V$--$I$)~}
\newcommand{\IVI}{$I$--($V$--$I$)~}
\newcommand{\VBV}{$V$--($B$--$V$)~}
\newcommand{\BV}{($B$--$V$)~}

\slugcomment{To appear in the Astronomical Journal, 2003 Dec issue}
\shorttitle{Sextans Dwarf Spheroidal Galaxy} \shortauthors{Lee et al.}

\begin{document}

\title{Deep Wide Field  $BVI$ CCD Photometry of the Sextans Dwarf Spheroidal Galaxy\altaffilmark{*}}
\author{Myung Goon Lee\altaffilmark{1,2}, Hong Soo Park$^1$, Jang-Hyun Park$^3$, Young-Jong Sohn$^4$,
Seung Joon Oh$^1$, In-Soo Yuk$^3$, Soo-Chang Rey$^4$, Sang-Gak Lee$^1$, Young-Wook Lee$^4$,
Ho-Il Kim$^3$, Wonyong Han$^3$,
Won-Kee Park$^1$, Joon Hyeop Lee$^1$, Young Beom Jeon$^3$, Sang Chul Kim$^{1,3}$ }

\affil{To appear in the Astronomical Journal, 2003 Dec issue}


\altaffiltext{1}{Astronomy Program, School of Earth and Environmental Sciences,
Seoul National University, Seoul 151-742, Korea, Email: mglee@astrog.snu.ac.kr }
\altaffiltext{2}{Visiting Investigator at the Department of Terrestrial Magnetism, Carnegie Institution of
Washington, 5241 Broad Branch Road, N.W., Washington, D. C. }
\altaffiltext{3}{Korea Astronomy Observatory, Daejeon 305-348, Korea}
\altaffiltext{4}{Center for Space Astrophysics, Yonsei University, Seoul 120-749, Korea}

\altaffiltext{5}{*Based on observations obtained at the Canada-France-Hawaii
Telescope (CFHT) which is operated by the National Research
Council of Canada, the Institut National des Science de l'Univers
of the Centre National de la Recherche Scientifique of France,
and the University of Hawaii.
This work is organized by the Korea Astronomy Observatory.}

\begin{abstract}
We present deep wide field  $VI$ CCD photometry of the Sextans dwarf spheroidal galaxy (dSph)
in the Local Group, covering a field of $42' \times 28'$ located at the center of
the galaxy (supplemented by short $B$ photometry).
The limiting magnitudes with 50 \% completeness are $V=24.4$ mag and $I=23.6$ mag.
Color-magnitude diagrams of the Sextans dSph show well-defined red giant branch (RGB),
blue horizontal branch (BHB), prominent red horizontal branch (RHB), asymptotic giant branch (AGB),
$\sim$120 variable star candidates including RR Lyraes and anomalous Cepheids,
$\sim$230 blue stragglers (BSs),
and main sequence (MS) stars.
The main sequence turn-off (MSTO) of old population is found to be located
at $V\approx 23.7$ mag and $(V-I)\approx 0.56$.
The distance to the galaxy is derived using the $I$-band magnitude of the
tip of the RGB at $I$(TRGB)$=15.95\pm0.04$:
$(m-M)_0=19.90\pm0.06$ for an adopted reddening of $E(B-V)=0.01$.
This estimate agrees well with the distance estimate based on the mean $V$-band magnitude
of the HB at $V$(HB)$=20.37\pm0.04$.
The mean metallicity of the RGB is estimated from the $(V-I)$ color: 
[Fe/H]$=-2.1\pm0.1$(statistical error) $\pm0.2$(standard calibration error) dex,
with  a dispersion of $\sigma$[Fe/H]=0.2 dex.
The age of the MSTO of the main old population is estimated to be similar to that of
the metal-poor Galactic globular cluster M92, and there are seen some stellar populations
with younger age.
There is found to be one RGB bump at $V=19.95\pm0.05$ mag ($M_V=0.03$ mag),
and a weak brighter bump at $V=19.35\pm0.05$ mag($M_V=-0.58$ mag) which is probably an AGB bump.
The $V$-band luminosity function of the RGB and MS stars is in general similar to that of
the globular cluster M92, with a slight excess of stars in the magnitude range brighter than
the MSTO with respect to that of M92.
The bright BSs are more centrally concentrated than the faint BSs.
The $V$-band luminosity function of the BSs in the inner region
is found to extend to a brighter magnitude  and to have a  flatter slope
compared with that of the BSs in the outer region.
Significant radial gradients are seen for several kinds of populations:
the RHB, the red RGB, the red subgiant branch (SGB), and the bright BSs are more
centrally concentrated toward the center of the galaxy, compared with
the BHB, the blue RGB, the blue SGB, and the faint BSs, respectively.

\end{abstract}

\section{Introduction}

Dwarf galaxies are believed to be basic building blocks of the large scale structures
in the universe. At the faintest end of the dwarf galaxies,
there are dwarf spheroidal galaxies (dSphs).
To date nine dSphs are known to be around our Galaxy.
These dSphs are the faintest galaxies among the known galaxies \citep{mat98, gre03}.
Some of these galaxies are similar in luminosity to the brightest globular clusters
in our Galaxy, but  much bigger in size than the latter,
making these galaxies distinguishable from globular clusters with similar masses.

These dSphs, due to their proximity and low crowding, provide us
with an excellent opportunity
to investigate in detail the various properties of stellar populations
and the star formation history in low mass stellar systems.
There have been many studies of these galaxies since the first discovery of this kind of
galaxy starting with the Sculptor and Fornax dSphs in 1938 \citep{sha38}.

Among these dSphs, the Sextans dSph is the least studied galaxy.
The Sextans dSph was discovered in 1990 by \citet{irw90} who used UKST sky survey plates.
It is the second last newcomer in the Galactic dSphs. (The last newcomer
is the Sagittarius dSph which was discovered in 1994 by \citet{iba94}.)
It has the lowest central surface brightness ($\mu_V(0)=26.2 \pm 0.5$ mag arcsec$^{-2}$)
among the known galaxies in the Local Group and
has a high mass to luminosity ratio ($M/L_V\approx 40 M_\odot/L_{\odot,V}$)
\citep{mat91,mat98}.
The Sextans dSph is large in the sky
(with core radius of $16.6\pm1.2$ arcmin and tidal radius of $160\pm50$ arcmin \citep{irw95})
so that it is not easy to study all the stellar populations
in the large fraction of the galaxy.
Previous photometric studies of this galaxy were based on either deep small-field photometry
or shallow wide field photometry \citep{irw90, mat91, mat95, bel01a}.
Recently \citet{dol02} presented a comparative study of star formation history
of seven dSphs in the Local Group
based on numerical modeling with the HST/WFPC2 photometry.
However, the Sextans dSph was not included in his study, because of no available HST/WFPC2 data
of the Sextans dSph.

In this paper we present a study of stellar populations in the Sextans dSph,
based on deep wide field $BVI$ CCD photometry which reaches below the main sequence turnoff and
covers a $42' \times 28'$ field including the center of the galaxy.
Star-formation history of the Sextans dSph based on the modeling with this photometry will be given separately
(Lee et al. (2003), in preparation). 

This paper is composed as follows.
We describe our observations and data reduction in Sec. 2, and present the color-magnitude
diagrams (CMDs) in Sec. 3. Variable star candidates are detected in Sec. 4,
and fiducial sequences in the CMD are derived in Sec. 5.
Distance and metallicity of the Sextans dSph are derived in Sec. 6 and 7.
Sec. 8 presents a study of bumps seen on the red giant branch (RGB), Sec. 9 shows the luminosity function
of the RGB and the main sequence, and Sec. 10 studies the properties of blue stragglers
in the Sextans dSph. Ages of stellar populations are approximately estimated in Sec. 11.
Sec. 12 presents a  study of radial gradient of several kinds of populations in the Sextans dSph.
Some issues related with the results are discussed in Section 13, and
primary results are summarized in the last section.


\section{Observations and Data Reduction}

\subsection{Observations}

$BVI$ CCD images of the Sextans dSph were obtained at the 3.6-m Canada-France-Hawaii
Telescope (CFHT) on the nights of UT 2001 February 16-17, using the CFH12K mosaic CCD camera.
The CFH12K camera is composed of twelve $2048 \times 4096$ pixel chips
($12,288 \times 8,192$ pixels in total), covering
a total field of $42\arcmin \times 28\arcmin$ on the sky at the f/4 prime focus of the CFHT.
The pixel scale is $0\farcs206$ per pixel.
One field covering the central region of the Sextans dSph was observed
on the clear nights of Feb 16 and 17, 2001.
Fig. 1 displays a finding chart for the observed CFH12K field of the Sextans dSph
on the digitized Palomar Observatory Sky Survey map.
Most of the bright stars seen in Fig. 1 are foreground stars, and
the stars of the Sextans dSph are barely visible.
The observation log is listed in Table 1.
Seeing ranged from 0.8 to 1.3 arcsec.
While both long and short exposures were taken with $V$ ($3 \times 1200$ sec and 60 sec) and
$I$ filters ($3 \times 600$ sec and 60 sec), only a short exposure was secured with
$B$ (60 sec).
Therefore long exposure images were used to get deep $VI$ photometry of the faint stars
and short exposure images were used to get $BVI$ photometry of bright stars and
to select variable star candidates.

%
In addition, $BVI$ CCD images of smaller fields overlapped partially with the CFH12K field
were obtained at the 1.8 m telescope of the Bohyunsan Optical Astronomy Observatory (BOAO)
in Korea, using a thinned SITe $2048\times 2048$ CCD camera during the observing
runs for 2001-2002.
The field of view of the BOAO CCD image is $11\farcm6\times11\farcm6$,
and the pixel scale is $0\farcs3438$ per pixel at the f/8 Cassegrain focus of the BOAO telescope.
Seeing ranged from 1.0 to 2.4 arcsec.
These BOAO CCD images were obtained with a purpose to calibrate the CFH12K photometry
of the Sextans dSph.
Five fields were observed, being overlapped partially with the CFH12K field.
The gain and readout noise of the BOAO CCD camera are 1.8e $^-/$ADU and 7.0 $e^-$, respectively.

%

\subsection{Data Reduction}

Raw CFH12K images were preprocessed using the FITS Large Image Processing Software (FLIPS).
The FLIPS is a highly automated software package developed for rapid reduction of the
large amount of data by Jean-Charles Cuillandre at the CFHT.
A detailed description of the FLIPS is given in \citet{kal01}.
Raw BOAO images were preprocessed using the IRAF.

Instrumental magnitudes of the objects in the CFH12K CCD images were derived using the point
spread function (PSF) fitting routine in the digital
photometry program DAOPHOT/ALLFRAME \citep{ste94}. Then we derived the stellarity
of the detected objects using the digital source extracting program SEXTRACTOR \citep{ber96}.
Stellarity is a very useful parameter to separate the point sources and the extended sources
in images.
Finally we selected as point sources the objects with stellarity larger than 0.3.
Instrumental magnitudes of the objects in the BOAO CCD images were derived using
the aperture photometry routine in DAOPHOT.

Instrumental magnitudes of the point sources in the BOAO CCD images were transformed
onto the standard Johnson-Cousins photometric system using the photometric standard
stars \citep{lan92} observed on the photometric night of Nov 19, 2001 at the BOAO.
An aperture of radius 7 arcsec as used in \citet{lan92} was used to derive the aperture
magnitudes of the standard stars in the BOAO CCD images.
The standard transformation equations are
$B=b+0.167 (b-v) -0.412X -0.903$,
$V=v-0.097 (b-v) -0.214X -0.717$,
$V=v-0.066 (v-i) -0.217X -0.752$, and
$I=i+ 0.026 (v-i) -0.139X -0.988$,
where the upper cases represent the standard magnitudes,
the lower cases the instrumental magnitudes (with a zero point of 25.0),
and $X$ the airmass.
The corresponding rms's are, respectively, 0.029, 0.015, 0.019 and 0.037
(see \citet{jhlee02} for details).
The aperture magnitudes of the bright stars in the Sextans were derived for the aperture
radius of 7 arcsec, consistently with the standard stars.
The apertures of  radius of the FWHM were used first to derived
the small aperture magnitudes, which were converted to the 7 arcsec radius aperture
magnitudes applying aperture correction. Then these magnitudes were transformed onto the
standard system using the equations above.

The instrumental magnitudes of the point sources in the CFH12K images were
transformed onto the standard system using the BOAO photometry of the bright stars 
in the fields overlapped with the CFH12K field of the Sextans dSph.
The mean color coefficients for transformation were derived from the average of
the color coefficients for 12 chips of the CFH12K camera:
$B=b+0.004(\pm0.044)(b-v)+zero(B)$,
$V=v-0.028(\pm0.048)(b-v)+zero(V_{bv})$,
$V=v-0.019(\pm0.046)(v-i)+zero(V_{vi})$, and
$I=i+0.010(\pm0.024)(v-i)+zero(I)$.
Using these mean color coefficients, we derived the zero points for each chip, which
are listed in Table 2 and are displayed in Fig. 2.
The rms's of the zero points we derived for each chip are 0.02$\sim$0.04 mag for $B$, and
0.01$\sim$0.03 mag for $V$ and $I$.

The array of the chip numbers shown in Fig.2 represents the real positions of the chips
in the camera.
Fig. 2 shows that the zero points for 12 chips in the CFH12K camera change only by
about 0.1 mag depending on the position of the chips. The trend that the zero points
increase at the chips in the outer region of the field appears to be due to
vignetting in the telescope.

The total number of stars in the CFH12K photometry is about 23,800, which
are listed in Table 3 (provided fully in an electronic table).
Astrometric solutions of each chip in the CFH12K images were derived using the bright stars
in the Guide Star  Catalog 2.2 (http://www-gsss.stsci.edu/gsc/gsc2/GSC2home.htm),
and were applied
to derive the equatorial coordinates (RA(2000) and Dec(2000)) of the stars in the images.
Fitting errors  (rms) in astrometric solutions are 0.14--0.19 arcsec.
The center of the galaxy is considered to be located at
RA(2000)$=10^h 13^m 03^s$, and Dec(2000)$=-01\arcdeg 36\arcmin 54\arcsec$ \citep{mat98}.

The mean errors of the CFH12K photometry are listed in Table 4 and are displayed
in Fig. 3. These errors are based on the internal DAOPHOT errors. In Fig. 3 we also
plot the external mean errors which are based on the difference between the input magnitudes
and the output magnitudes of the artificial stars which will be described in the following.
Both agree well except the fact that at the faint end
the mean internal error is slightly larger than the external error.
We used the internal mean errors in this study.

%
%
%
%
%
%
%
%
\subsection{Photometric Completeness}

We have derived completeness of our CFH12K $VI$ photometry using the artificial star
experiment with the ADDSTAR routine in DAOPHOT. We performed this artificial star
experiment for one chip (chip 8) in the center to derive a typical value.
Photometric completeness appears to change little among the chips,
because the zero points of the chips change only at the level of about 0.1 mag,
and because the degree of crowding affecting the stellar photometry is very low
over the entire field covered by the CFH12K camera.

We created the artificial stars so that the colors, magnitudes, luminosity functions
of those stars are similar to those of the observed stars in the Sextans dSph.
We added 2000 artificial stars with $19.8<V<25.8$ mag to each CCD image, and the total number
of added stars in five pairs of $V$ and $I$ images is 10,000.
We applied the same procedures to the resulting images with artificial stars
as used originally in deriving the $VI$ photometry from the CCD images of the Sextans dSph.
Finally we derived the completeness as the ratio of the number of recovered stars
to that of the added stars ($\Lambda =$ N(recovered)/N(added)), which is listed in Table 5
and is plotted in Fig. 4.
Fig. 4 shows that our photometry is 50 \% complete at $V= 24.4$ mag and $I= 23.6$ mag.

%
%
%
%

\subsection{Comparison with Previous Photometry}

\citet{mat95} presented $BV$ photometry of a smaller field ($18'\times 18'$)
in the Sextans dSph than ours.
We have compared our photometry with theirs for about 900 non-variable stars common between
the two, the results of which are displayed in Fig. 5.
During the comparison process, it is found that there are slight systematic positional
differences in the coordinates between the two studies: $\Delta$RA$ \approx +1.3$ arcsec
and $\Delta$Dec$ \approx +0.6$ arcsec where $\Delta$ means this study minus
\citet{mat95}. These differences are larger than our astrometric errors (0.14--0.19 arcsec)
as described in Sec. 2.2.

Fig. 5(a) shows that the difference in $V$ ($\Delta V$(This study--Mateo et al. (1995))
increases slightly as the magnitude increase.
A linear fit with two-sigma clipping to the data for 651 stars with $V<22$ mag yields
$\Delta V = 0.0278 V-0.5494$ (rms=0.0099).
On the other hand, the difference in $(B-V)$ in Fig. 5(b) is on average almost zero.
A linear fit with two-sigma clipping to the data for 653 stars with $V<22$ mag yields
$\Delta (B-V) = -0.0036 V + 0.0490$ (rms=0.0977).
The source of the difference in $V$ is not clear.

%
%

\section{Color-Magnitude Diagrams}

Fig. 6(a) displays a \VVI diagram of the measured stars in the CFH12K images.
There are included some foreground stars and background objects in this figure.
We need a control field to see the effect of these foreground and background objects.
The size of our CFH12K field ($42\arcmin \times 28\arcmin$) is, although wide, still smaller
than the size of the Sextans  which has a core radius of $16.6\pm1.2$ arcmin
and a tidal radius of $160\pm50$ arcmin \citep{irw95, gre02}.
There is no suitable region which can be considered as a control field
inside our CFH12K field.

Therefore we display, in Fig. 6(b),
a color-magnitude diagram (CMD) of a field close to Palomar 3,
a Galactic globular cluster, which is about 3 degrees from the Sextans dSph
in the sky, given by \citet{soh03}.
The $VI$ photometry of this field were obtained using the same instrument during
the same observing run as the Sextans dSph was observed (on Feb 16, 2001).
The exposure times are 1500 sec for $V$ and 1200 sec for $I$.
Stars brighter than $V=17$ mag were saturated in the control field images.
This control field is of the same angular size as that of the Sextans dSph field,
 and is centered at
 RA(2000)$=10^h 04^m 59^s.4$ and Dec(2000)$ = -00\arcdeg 28\arcmin 43\arcsec$,
 about 33 arcmin south from Palomar 3.
\citet{soh03} used only DAOPHOT to derive
 the photometry of this control field, while we used DAOPHOT/ALLFRAME and  SEXTRACTOR to derive
 the photometry of the Sextans dSph.
So the information of the SEXTRACTOR stellarity of the control field is not available.
There may be some variation of the field stars over 3 degrees in the sky so that this control
field can be used only a guide.

Comparison of the two CMDs in Fig. 6 shows that
 most bright yellow and red stars with $(V-I)>0.6$ seen uniformly in the CMD are considered
to be foreground stars,
and a large number of faint red objects with $V>23$ mag are compact background galaxies.
Note that most of the background galaxies were subtracted mostly
in the CMD of the Sextans dSph where the stellarity of the SEXTRACTOR  was used
for selecting point sources (the stellarity of the SEXTRACTOR was not used for deriving
the photometry of the control field).
%
%

The CMD of the Sextans dSph in Fig. 6(a) shows several notable features.
(1) There is seen a steep and well-defined red giant branch (RGB) extending up to
$V \approx 17.4 $ mag at $(V-I)\approx 1.35$.
(2) There is a prominent red horizontal branch (RHB) at
$V\approx 20.35$ mag and $(V-I)\approx 0.7$, and a weaker
blue horizontal branch (BHB) at $(V-I)\approx 0.2$ with a tail extending down
to $V\approx 20.9$ mag at $(V-I)\approx 0.05$.
(3) There are seen a small number of stars at the blue side of the RGB above the
HB, which are AGB stars.
(4) There are some stars between the BHB and the RHB, which are mostly
RR Lyrae variable stars, as shown in the following section.
(5) There are seen many stars below $V\approx 23$ mag which are main sequence (MS)
stars with old age.
A main sequence turn-off (MSTO) is seen at $V \approx 23.7 $ mag at $(V-I)\approx 0.56$.
A large spread in color at the faint magnitude is mostly due to photometric errors
(as listed in Table 4).
(6) There is a distinguishable group of blue stragglers extending up to
$V\approx 21.2$ mag at $(V-I)\approx 0.0$ above the MS.

\section{Variable Star Candidates}

It is expected that there exist many variable stars in the CFH12K field of the Sextans dSph,
which are included in the CMD in Fig. 6(a).
\citet{mat95} found 44 variable stars in a total area of $18'\times 18'$ in the
Sextans dSph: 36 RR Lyrae, 6 anomalous Cepheids, one long-period red variable,
and one foreground contact binary.
Our CFH12K field is 3.6 times larger than the field used for the variable search by \citet{mat95}.
Since long exposure CFH12K images and short exposure CFH12K images were obtained
with a time lag of about one day, they are useful to identify variable star candidates.

We have searched for variable star candidates by comparing the photometry of
short $VI$ exposures and long $VI$ exposures. Fig. 7 shows the difference in
$V$ magnitudes between the short and long exposures.
We consider, as variable star candidates, 122 stars (marked by the squares in Fig. 7)
with magnitude differences larger than three times the mean error at given magnitudes,
which are listed in Table 6.  Considering that the amplitude of variability
is larger in $V$ than in $I$, we selected, as the variable star candidates, those which
were detected in both $V$ and $I$ or in $V$ only.
Eighty five among these 122 candidates were detected as variable
stars in both $V$ and $I$, and thirty seven were in $V$ only.

Table 6 lists the $BVI$ photometry of the variable star candidates
which are derived from the short exposure images.
Most of the stars in the vertical structure with $19.5<V<20.8$ mag
seen in Fig. 7 are probably RR Lyrae stars.
Among the 43 variable stars the positions of which were given by \citet{mat95},
23 were recovered  both in $V$ and $I$, 7 were in $V$ only,
2 were in $I$ only, and 11 were not recovered.
Among the 35 RR Lyraes with positions given by \citet{mat95}, 26 RR Lyrae were recovered,
and 9 RR Lyrae were not recovered in our search
(one of them was recovered in $I$ only so that it was considered as not-recovered).
Therefore the recovery rate of RR Lyrae in our
search based on two-epoch photometry is estimated to be 74 \%.

%
%
%
%

Fig. 8 shows the \VVI and \VBV diagrams of these variable star candidates
(open squares).
The RR Lyrae stars and anomalous Cepheids found  by \citet{mat95} are also plotted in Fig. 8
(red and blue triangles, respectively).
There are seen two main groups of variable star candidates which are
probably the members of the Sextans dSph, as marked in Fig. 8(a).

First, the RR Lyrae Group (labeled by RR):
Most of the variable star candidates as well as the previously known RR Lyraes
are located in the RR Lyrae instability region.
There are 90 RR Lyrae candidates including known RR Lyraes.
There are three new RR Lyrae candidates in the field used for the variable star search by
\citet{mat95}. These three RR Lyrae candidates may not be real RR Lyrae variables or might have been
missed by \citet{mat95}.
Since the latter is unlikely, we estimate that our false detection rate of RR Lyraes is 9 \%.
Since the recovery rate of
our search for RR Lyraes is 74 \% (as shown above),
the total number of the RR Lyraes in the CFH12K field is estimated to be
(90 \time 0.91)/0.74 = 111 . 
Note that the RR Lyrae candidates are located along the region crossing the HB with some angle.
This is because the photometry of these
variable star candidates are based on the single-epoch measurement, not on the mean
magnitudes over the period.

Secondly, the anomalous Cepheid Group (labeled by AC):
A small number of variable star candidates above the HB may be
anomalous Cepheids. There are a few candidates in addition to the known
six anomalous Cepheids. However, they are overlapped with the brightest RR Lyraes so that
 time-series photometry is needed to identify them.

Two of the faint variable star candidates are located
in the blue straggler region in Fig. 8(a).
These two stars may be SX Phoenicis stars or eclipsing binaries.
SX Phoenicis stars are Population II short-period
pulsating variables found in Galactic globular clusters and dSphs \citep{rod00},
and all known SX Phoenicis stars are found to be  blue stragglers \citep{jeo03}.
However, these two stars were detected as variable only in $V$ so that it needs
further confirmation.
Other variable star candidates not belonging to the above groups are probably
foreground variable stars in our Galaxy or other kinds of variable stars in the Sextans dSph.

Fig. 9 displays a finding chart of these variable star candidates.
Time-series photometry  obtaining the full light curves of these variable star candidates
is needed to study in detail their properties.
In this study we use them only for identifying the variable stars
in the color-magnitude diagrams, by which we can distinguish the member stars of
the Sextans dSph from the foreground stars and we can count the number of
variable stars in the comparison of stellar populations.
%
%

\section{$VI$ Fiducial Sequences}

We have derived approximate fiducial sequences in the \IVI diagram of the Sextans dSph
by using eye-estimation, the mean color, and the number density
in the color magnitude diagram of the entire CFH12K field.
These fiducial sequences are listed in Table 7, and are displayed by the solid lines in the
\VVI diagram of Fig. 10. Variable stars (and candidates) are not plotted in Fig. 10.

In Fig. 10 we also display the number density contour map for the MS
stars with $V>23$ mag which was used as an aid in deriving the fiducial sequence of the MS.
Two dashed lines at the left and right of the RGB fiducial line represent
the boundary with $\Delta(V-I)=\pm0.1$ mag.
Mean photometric errors versus magnitudes are also plotted by the error bars in the leftside
of the figure.
While the color spread in the bright RGB is much larger than the mean photometric errors,
the color spread in the SGB and MS is getting comparable to the mean
photometric errors.
The fiducial line for the MS shows that the bluest part of the MS is located
at $(V-I)\approx 0.56$ and $V\approx 23.7$ mag.

%
%

\section{Distance}

Reddening, distance, and metallicity are derived iteratively from the photometry.
Here we describe how we adopt the reddening and how we estimate the distance to the Sextans.
Metallicity will be derived  in the following section.

\subsection{Reddening}

The position of the Sextans dSph in the sky ($l=243^{\arcdeg}.498$ and $b=42^{\arcdeg}.27$)
indicates that the foreground reddening toward this galaxy is low and
 the morphological type of this galaxy implies that the internal
 reddening is negligible.
While the old galactic extinction map \citep{bur82} gives a value
for the foreground reddening toward the Sextans dSph, $E(B-V)=0.02$ ,
a new galactic extinction map \citep{sch98} gives
 a slightly higher value, $E(B-V)=0.047$.
\citet{mat98} lists $E(B-V)=0.03\pm0.01$ for the Sextans dSph in the review of
the dwarf galaxies of the Local Group.

On the other hand, \citet{mat95} estimated the reddening directly from the
$BV$ photometry of RR Lyrae stars,
using the fact that the minimum light \BV colors of RR$ab$ stars are constant
\citep{pre64}. They used a relation between the color, period and metallicity:
$(B-V)_{0,min} = 0.362 +0.24 \log P + 0.056$ [Fe/H] (their eq. (5)).
Assuming [Fe/H]$=-1.6$ dex for the short period variable stars they found in this galaxy,
they derived a reddening value $E(B-V)=0.037\pm0.027$.
If [Fe/H]$=-2.1$ dex (as derived in Sec. 7) is adopted instead,
the reddening value will be decreased to $E(B-V)=0.009$.
Our $(B-V)$ color is 0.024 bluer than that of \citet{mat95} at the mean magnitude of
RR Lyrae stars $V=23.4$ mag, according to the equation in Sec. 2.4.
If we consider this difference, then the reddening value will be negative.
In this study we adopt the reddening of $E(B-V)=0.01\pm0.02$ for the Sextans dSph.

\subsection{Distance}

We have estimated the distance to the Sextans dSph using the $I$-band magnitude of the tip
of the RGB (TRGB), following the method described in \citet{lee93b}.
The TRGB method, which was systematically applied to nearby galaxies for the first
time by \citet{lee93a} and \citet{lee93b}, has proved to be an excellent distance
indicator for resolved galaxies with old giants (for example, see \citet{kim02, lee02}).
This method takes advantage of the fact that $I$-band magnitude of the TRGB is little sensitive
to metallicity (for [Fe/H]$<-0.7$ dex) or age (for older than a few Gyr), and can
be used efficiently to estimate the distances to resolved galaxies with old populations.

The $I$ magnitude of the TRGB is estimated using the $I-(V-I)$ diagram
in Fig. 11 and the luminosity function of red giant stars.
We have derived the $I$-band luminosity function of red giant stars by counting
stars around the RGB in the \IVI diagram
(stars located inside the RGB boundaries shown in Fig. 10), displaying it in Fig. 12.
There may be included a small number of field stars in this luminosity function, but
the CMDs in Fig 6(a) and 6(b) show that the their contribution to this luminosity
function is negligible.

Fig. 12 shows that,
as the magnitude increases, there is a sudden
increase at $I=15.95\pm0.05$ mag in the luminosity function,
which corresponds to the TRGB seen in the color-magnitude diagram in Fig. 11.
The mean magnitude and color of the TRGB  are derived
 from the mean of six bright red giants with $15.88<I<16.0$ mag:
$I=15.95\pm0.04$ and $(V-I)(TRGB)=1.36\pm0.04$.

The bolometric magnitude of the TRGB is then calculated from
$M_{\rm bol}=-0.19{\rm [Fe/H]} - 3.81$ \citep{dac90}.
Adopting a value for the metallicity of [Fe/H]  $= -2.1 \pm 0.1$  dex as
estimated in the following section, we obtain a value for
the bolometric magnitude of $M_{\rm bol}=-3.42$ mag.
The bolometric correction at $I$ for the TRGB is estimated
to be BC$_I=0.55$ mag,
adopting a formula for the bolometric correction
BC$_I$ = 0.881 -- 0.243$(V-I)_{\rm TRGB}$ \citep{dac90}.
The intrinsic $I$ magnitude of the TRGB is then given by
$M_I= M_{\rm bol} - {\rm BC}_I = -3.97$ mag.
Finally the distance modulus of the Sextans dSph is obtained:
$(m-M)_0 = 19.90 \pm 0.06$ mag
(corresponding to a distance of $95.5\pm 2.5$ kpc)
for an adopted extinction of $A_I=0.02$ mag.

%
%
%
%

We have derived the distance to the Sextans dSph also  using the mean $V$-band magnitude
of the HB stars. The mean $V$-band magnitude of the RHB is derived from
the 87 stars with $20.2<V<20.5$ mag and $0.66 <(V-I)<0.83$:
$<V(RHB)>=20.35\pm0.04$.
This is very similar to the mean magnitudes of the RR Lyrae variable candidates
and BHB:
$<V(RR)>=20.38\pm0.19$ for 77 stars with $20.2<V<20.5$ mag and $0.26 <(V-I)<0.68$ ,
 and
$<V(BHB)>=20.38\pm0.04$ for 19 stars with $20.2<V<20.9$ mag and $0.12 <(V-I)<0.24$
(excluding the blue-tail stars).
A mean of these three values is derived to be $20.37\pm 0.04$.
Using $M_V(RR)=0.17{\rm [Fe/H]}+0.82$ \citep{lee90} and the mean magnitude of the HB, 20.37 mag,
we derive a distance modulus: $(m-M)_0(HB)=19.89\pm0.04$ for an extinction of $A_V=0.03$.
This value is in excellent agreement with the estimate based on the TRGB above.

Our estimates of the distance to the Sextans dSph are slightly larger than the previous estimates:
\citet{mat95} gave $(m-M)_0=19.67\pm0.15$ ($d=86\pm6$ kpc) for $E(B-V)=0.037\pm0.027$, and
\citet{mat98} listed $(m-M)_0=19.67\pm0.08$ ($d=86\pm4$ kpc) for $E(B-V)=0.03\pm0.01$.
This difference appears to be due to combination of photometry difference, reddening, and
the distance modulus of M15 which was used a  distance reference by \citet{mat95}.
Our $V$ magnitude is 0.02 mag fainter than that of \citet{mat95} for the mean magnitude
of the $HB$ stars, according to the equation in Sec. 2.4.
In deriving the distance to the Sextans dSph \citet{mat95} assumed a distance modulus
of $(m-M)_0 =15.03$ for M15, while
recent references list slightly larger values: $(m-M)_0=15.06$ \citep{har96},
and 15.12 \citep{fer99}.
Considering all these, their distance modulus of the Sextans will be
$(m-M)_0=(19.80-19.86) \pm0.15$ which is
in better agreement with ours.

\section{Metallicity}

We have estimated the mean metallicity of the RGB stars in the Sextans dSph
using  the $(V-I)$ color of the stars 0.5 mag fainter than the TRGB, $(V-I)_{-3.5}$.
This color is measured from the mean value of the observed
colors of 16 red giant branch stars  with $16.2<I<16.7$ mag
to be 
$(V-I)_{-3.5}=1.26\pm0.01$(m.e.) with a dispersion of 0.05.
The reddening-corrected color is $(V-I)_{-3.5,0} = 1.25$ for the adopted reddening.
Using the relation [Fe/H]$=-12.64+12.6(V-I)_{3.5,0} -3.3 (V-I)^2_{-3.5,0}$ \citep{lee93b},
we derive 
a value for the mean metallicity of [Fe/H] $= -2.1 \pm 0.1$ dex, 
with a metallicity dispersion of 0.2 dex.
This estimate is also consistent with the estimate based on
the mean color of the RGB at $M_I=-3.0$ mag.
The mean color of the RGB at $M_I=-3.0$ mag (1.0 mag below the TRGB) is derived
from the mean of the colors of 25 red giants with $16.7<I<17.24$ mag
to be $(V-I)_{-3.0}=1.16\pm0.01$ with a dispersion of 0.04. 
This yields an estimate of [Fe/H]$=-2.1\pm0.1$, the same as above.
The error in the standard calibration, $e(V-I)=0.04$ leads to an error of the mean metallicity,
0.2 dex.
Finally the mean metallicity is derived to be
[Fe/H] $= -2.1 \pm 0.1$(statistical mean error)$\pm0.2$(calibration error) dex.

In Fig. 11, we compare the \IVI diagram of the Sextans dSph with the fiducial sequences
of metal-poor Galactic globular clusters, M92, M15, NGC 6397, and M3 with different metallicities.
The sources for the fiducial sequences of the globular clusters are
\citet{joh98, khlee03} for M92, \citet{dac90} for M15 and NGC 6397,  and
\citet{joh98} for M3. In the case of M92, we have derived the fiducial sequence using the
$VI$ photometry of M92 given by \citet{khlee03}.
The metallicities of the globular clusters, M92, M15, NGC 6397, and M3, are
[Fe/H] = --2.24, --2.17, --1.91 and --1.66 dex, respectively \citep{har96, fer99}.
The distance moduli and reddening adopted for these clusters are:
$(m-M)_0=14.67$ and $E(B-V)=0.02 $ for M92,
$(m-M)_0=15.12$ and $E(B-V)=0.09 $ for M15,
$(m-M)_0=14.67$ and $E(B-V)=0.02 $ for NGC 6397, and
$(m-M)_0=14.99$ and $E(B-V)=0.01 $ for M3 \citep{har96, fer99}.

In Fig. 11 it is seen that the mean RGB of the Sextans  agrees well with those
of M92 and M15, and that the red boundary of the upper RGB of the Sextans is close to the
fiducial sequence of NGC 6397 (with [Fe/H] $= -1.91$ dex).
This shows that the metallicity of most RGB stars is lower than [Fe/H] $= -1.9$ dex.
The positions of the BHB and RHB of the Sextans are consistent with the fiducial lines of M92 (BHB) and
M3 (BHB and RHB), although the RGB of the Sextans is bluer than that of M3 (with [Fe/H]$=-1.66$ dex).
A small number of stars with $16.5<I<19.5$ mag slightly bluer
than the bright RGB sequence of M15 are mostly AGB stars.
The AGB stars of the Sextans dSph are located well along the
AGB sequence of M3.

We have derived the $(V-I)$ color distribution of the red giant stars as a function of magnitude,
displaying it in Fig. 13.
In Fig. 13 $\Delta(V-I)$ represents the color difference with respect to the color of the fiducial
line of the Sextans dSph.
Figs. 13(a) and 13(b) shows that the color spread of the bright stars at $15.9<I<18$ mag
(just below the TRGB) is significant, ranging from $\Delta (V-I)\approx -0.08$ to  $\approx 0.05$,
with most of stars at $-0.04<\Delta (V-I)<0.03$.
The color distribution at $-0.04<\Delta (V-I)<0.03$ is roughly uniform, although the number of
stars is small.
The color spreads of the  bright stars ($I<20$ mag) in Fig. 13 are much larger
than the mean photometric errors (marked by the horizontal bar at above $\Delta(V-I)=0$),
and they get close to the mean errors at the faint magnitude, as shown also in Fig. 10.

The color distribution of the stars at $18<I<19$ mag in Fig. 13(c) shows clearly two components:
one major component centered at $\Delta (V-I)=0$, and one weaker component centered
at $\Delta (V-I)=-0.05$. These two components correspond, respectively, to the RGB and the
AGB, as also shown in the CMD in Figs. 10 and 11.
Gaussian fitting to these data, as shown by the dashed lines and the solid line in Fig. 13, yields
one Gaussian with center at $\Delta (V-I)=0.00 $ and width of 0.04 for the RGB component,
and another Gaussian with center at $\Delta (V-I)=-0.05 $ and width of 0.06 for the AGB component.
This indicates that a small number of bright stars with $15.9<I<18$ mag at $-0.08<\Delta (V-I)<-0.04$
(at the blue side of the main group), or at lease a part of them, are probably AGB stars.
Therefore the color spread due to the group of the RGB is considered to be
mainly $-0.04<\Delta (V-I)<0.03$, with a small number of stars outside this range.
This large color spread of the upper RGB at $15.9<I<17$ mag ($-0.04<\Delta (V-I)<0.03$)
is considered to be mainly due to the metallicity spread in the RGB component,
as large as $\Delta$[Fe/H]=0.2 dex, as measured above (see also the arrow marks in Fig.13(a)).

There are several estimates of the metallicity of the stars in the Sextans dSph,
starting in 1991, which are summarized in Table 8.
These metallicity estimates were derived from the photometry of a large number of red giants,
low-resolution spectroscopy, or high resolution spectroscopy of a small number of red giants.
Table 8 shows that there is a significant difference among the mean values of the metallicity
([Fe/H]=--1.6 to --2.1 dex), while the metallicity dispersion estimates agree well
($\Delta$[Fe/H]$\approx 0.2$ dex).

Interestingly the distribution of previous estimates of the mean metallicity is roughly bimodal,
with one peak at [Fe/H]$\sim-1.6$ dex \citep{mat91, dac91, mat95, mat98},
and another peak at [Fe/H]$\sim -2.05 $ dex \citep{sun93,gei96,she01}.
This trend is seen in both the photometric estimates and spectroscopic estimates.
\citet{dac91} derived the metallicity of six giants in the Sextans from the low resolution
spectroscopy, obtaining [Fe/H]$=-1.7\pm0.25$ dex.
On the other hand, \citet{sun93} derived the metallicity of 43 giants in the Sextans dSph
(seven times larger than the sample used by \citet{dac91}) from the low
resolution spectroscopy, obtaining a mean value of [Fe/H]$=-2.05\pm0.04$ dex with a dispersion
of $0.19\pm0.02$ dex. The metallicity distribution they obtained shows a large range of
$-2.5<$[Fe/H]$<-1.5$ dex, following approximately a single Gaussian (see their Fig. 9).

Recently \citet{sun93}'s  result was supported by the high resolution spectroscopic data,
although the number of the sample is small, given by \citet{she01}.
\citet{she01} measured the metallicity of five bright red giants in the Sextans dSph
using the high resolution spectroscopy. They found that the metallicity of these stars range from
[Fe/H]= $-1.45\pm0.12$ dex to $-2.85\pm0.13$ dex, and that the weighted mean metallicity is
$-2.07\pm0.10$ dex with a dispertion of 0.21 dex.
In addition, they found that the average even-Z abundance of these stars in the Sextans dSph,
[$\alpha$/Fe]$=0.02\pm0.07$ dex, is significantly lower than that of the halo field stars
([$\alpha$/Fe]$=0.28\pm0.02$ dex) and the stars in M92 ([$\alpha$/Fe]$=0.34\pm0.04$ dex), although
these three samples have similar metallicity.

The results given by \citet{sun93} and \citet{she01} are considered to have more weights than
the result given by \citet{dac91}, because the former are based either on much larger sample or
on higher resolution spectroscopy compared to the latter.
Our mean metallicity estimate is very similar to the [Fe/H]$\sim -2.1$ dex peak,
and our metallicity spread value is consistent with previous estimates.
If the metallicity of the Sextans is [Fe/H]$=-1.7$ dex, $(V-I)$ color at $I\approx 16.5$ mag
of the RGB should be redder by about 0.1 than our measurement, as seen by the fiducial sequence of
M3 (with [Fe/H] $= -1.66$ dex) in Fig. 11, which is very unlikely.
In summary, the mean metallicity of the giant stars in the Sextans is estimated to be
[Fe/H]$= -2.1\pm0.2$ dex with a dispersion of 0.2 dex.

\section{Red Giant Branch Bump}

The RGB bump is a concentration of stars seen on the RGB around the magnitude level of the HB,
revealed by an excess in the differential luminosity function
or by a change of slope in the cumulative luminosity function \citep{fus90, fer99}.
Theoretically the RGB bump is supposed to appear where stellar evolution
in the first stage of the red giants is relatively slow,  when the H-burning shell
crosses the chemical discontinuity left over by the convective envelope
soon after the first dredge-up \citep{tho67, ibe68}.
The RGB bump has been detected in numerous Galactic globular clusters \citep{pio02} and
a few Local Group dwarf galaxies (Sculptor, Sagittarius, Sextans, Ursa Minor and Leo II dSphs) since
its first detection in 47 Tuc by \citet{kin84}
(see the references in \citet{cas97, fer99, mon02, bel02a, rie03, bel03} ).
The most dramatic example of the RGB bump detected thus far is
that in the Sagittarius dSph \citep{mon02}.

We have searched for the RGB bump in the Sextans dSph.
Fig. 14(a) shows $V$-band and $I$-band luminosity functions of the red giant stars within
the $\Delta (V-I)=\pm0.04$ color boundary covering most of the RGB stars,
as shown in the color distribution in Fig. 13.
Here we used a very narrow range of $\Delta (V-I)=\pm0.04$ color boundary
to avoid the contamination due to AGB stars as much as possible.
In Fig. 14(a) we display also the $V$-band luminosity function of the AGB stars with
$-0.10 < \Delta (V-I) < -0.04$.
The range of color of the AGB stars was determined from the color distribution
of stars with $18<V<19$ mag in Fig. 13(c).
In addition, we divided the RGB stars into two groups:  the blue RGB ($-0.04<\Delta (V-I)<0.00$)
and the red RGB ($0<\Delta (V-I)<0.04$).
$V$-band luminosity functions of these two groups were plotted in Fig. 14(a). 
While the differential luminosity functions were plotted in Fig. 14(a),
the $V$-band cumulative luminosity functions were plotted in Fig. 14(b).

In Fig. 14(a) both $V$-band and $I$-band differential luminosity functions
(black and magenta histograms, respectively)
show that there is a distinguishable peak at $V=19.95\pm0.05$ mag
(and $I=18.95\pm0.05$ mag), and
a weaker and brighter peak at $V=19.35\pm0.05$ mag (and $I=18.35\pm0.05$ mag).
The fainter peak (labeled as bump 1) is seen  clearly both in the blue and red RGB
(blue and red lines, respectively),
while the brighter peak (labeled as bump 2) is seen only in the blue RGB.

In addition, we also checked the presence of these peaks using the cumulative luminosity function
in Fig. 14(b).
In Fig. 14(b) the cumulative $V$-band luminosity functions
show changes of slopes at the same peak positions as shown in Fig. 14(a),
confirming that there exist two peaks
in the entire RGB and the blue RGB, and one peak in the red RGB.
These faint and bright peaks represent possibly RGB bumps.

Our results on these bumps 
are consistent with those based on the $B$-band luminosity function given by \citet{bel01a}.
\citet{bel01a} found a fainter bump (bump 1) at $B=20.8\pm0.1$ mag
(corresponding to $V\approx 19.95$ mag) and a brighter bump (bump 2) at $B=20.2\pm0.1$ mag
(corresponding to $V\approx 19.35$ mag).
They considered these two bumps as RGB bumps.
There is no doubt that the fainter bump is an RGB bump.
However, we argue below {\it that the brighter bump is an AGB bump, instead of an RGB bump}.

(1) The relation of the bump 2 and the AGB: In Fig. 14(a),
the luminosity function of the AGB stars (dashed line) shows
that most of the AGB stars are concentrated in the magnitude range of
$19.0<V<19.7$ mag,  showing two peaks of similar size,
at $V=19.15\pm0.05$ mag and $V=19.55\pm0.05$ mag.
The mean of these two magnitudes, $V=19.35$ mag, corresponds exactly to the magnitude of the bump 2.
The bump 2 is seen only in the blue RGB, not in the red RGB.
Therefore the bump 2 is an extension of the AGB in the CMD.

(2) The luminosity of the bump 2: The absolute magnitudes of the bumps 1 and 2 are, respectively,
$M_V=0.02$ mag and --0.58 mag for the distance derived in this study.
The luminosity of the RGB bump depends primarily on the metallicity
(in the sense that the RGB bump gets brighter as the metallicity decreases),
and weakly on the age.
The difference in $V$-band magnitude between the HB and the RGB bump $\Delta V^{bump}_{HB}$
is a useful metallicity indicator
which is free of distance and extinction \citep{fer99}.
We have derived $\Delta V^{bump}_{HB}=-0.45\pm 0.06$ mag for the bump 1, and
$-1.00\pm0.06$ mag for the bump 2.
$\Delta V^{bump}_{HB}$ of the bump 1 is within the range of the values for
the RGB bumps in the Galactic globular clusters ($\Delta V^{bump}_{HB} = -0.7$ to 0.8 mag),
but $\Delta V^{bump}_{HB}$  of the bump 2 is smaller than the minimum value
of the RGB bumps of the Galactic globular clusters \citep{fer99,rie03}.
The calibration between $\Delta V^{bump}_{HB}$ and [Fe/H] based on
the Galactic globular clusters is given by \citet{fer99} (their equation 6.2):
$\Delta V^{bump}_{ZAHB}$ = 0.67 [Fe/H] + 0.827, where [Fe/H] is given
in terms of \citet{zin85} scale as used in this study.
$V_{ZAHB}$ represents the magnitude of the zero-age HB, and is given
by $V_{ZAHB}=<V_{HB}>+0.106[{\rm M/H}]^2+0.236[{\rm M/H}]+0.193$  where [M/H] is
a global metallicity ([M/H] = [Fe/H] + [$\alpha$/Fe]) \citep{fer99}.
Using this calibration ([Fe/H] $= 1.49 \Delta V^{bump}_{ZAHB} - 1.234$),
the metallicity of the bump 1 is derived to be [Fe/H] $= -2.0 \pm0.1 $ dex.
This value is in good agreement
 with the mean metallicity derived from the color the RGB stars above.
The value of the $\Delta V^{bump}_{HB}$ for the bump 2
is outside the calibration range of the equation above, indicating that
[Fe/H]$<-2.4$ dex (if extrapolated, the metallicity will be [Fe/H]=--2.7 dex),
if it is an RGB bump.
However, there is little evidence for the existence of a significant amount of
this metal-poor population in the Sextans dSph, as shown in the color distribution
in Fig. 13.

From the above two evidences (that the bump 2 is an extension of the AGB, and that the bump 2 is
brighter than the brightest RGB bump in the Galactic globular clusters), we conclude
that the bump 2 is not an RGB bump, but an AGB bump. Therefore the Sextans dSph is found to have
one RGB bump and one AGB bump.

The AGB bump appears when the core He burning switches to the shell He burning.
AGB bumps are in general difficult to detect, because the number of stars in the AGB bump is
small due to short evolution time for the AGB bump.
AGB bumps have been detected in several Galactic globular clusters, and in some
galaxies in the Local Group \citep{gal98}.
Theoretical AGB bumps get brighter with increasing stellar mass and with
decreasing metallicity (see Fig. 15 in \citet{fer99}).
The calibration for $\Delta V^{AGB-bump}_{HB}$ and the metallicity [Fe/H]
is given by \citet{fer99}: $\Delta V^{AGB-bump}_{HB} = -0.16$[Fe/H]$_{CG97}-1.19$
([Fe/H]$_{CG97} = 6.25 \Delta V^{AGB-bump}_{HB} - 7.74$)
where $\Delta V^{AGB-bump}_{HB}$
represents the $V$ magnitude difference between the AGB bump and the HB.
Here [Fe/H]$_{CG97}$ represents the metallicity scale given by \citet{car97} (CG97 scale).
The CG97 scale yields about 0.2 dex higher metallicity than the Zinn scale for low to intermediate
metallicity globular clusters \citep{fer99}.
Using this calibration, the metallicity of this bump is derived to
be [Fe/H]$_{CG97} = -2.11 \pm 0.43$ dex (note that the error is large, because of the steep dependence
of the metallicity on $\Delta V^{AGB-bump}_{HB}$).
This value is  in agreement
 with the mean metallicity derived from the color of the RGB stars obtained in this study,
 supporting that this bump is an AGB bump.

This situation is very similar to the case of the Sculptor dSph for which \citet{maj99} found
two bumps with $\Delta V^{bump}_{HB}=-0.35$ and --0.9 (only 0.1 mag smaller than those
of the Sextans dSph).
\citet{maj99} derived the luminosity functions of the blue RGB ($-0.125<\Delta (B-V)<0.0$) and
red RGB $0.0<\Delta (B-V)<0.125$) of the Sculptor dSph,
finding that a brighter bump is seen on in the blue RGB
and a fainter bump is seen only in the red RGB (see their Fig. 3).
They considered these two bumps as the RGB  bumps.
However, the CMD of the Sculptor looks different from that of the Sextans in (a) that the
reddest part of the BHB is brighter than the bluest part of the RHB
and (b) the RGB of the Sculptor looks thicker than that of the Sextans dSph.


Recently  several studies came out reporting the detection of the RGB bumps in the dSphs around
our Galaxy.
\citet{bel02a} found from wide field $VI$ photometry that the Ursa Minor dSph has
a single RGB bump
at $V=19.40\pm0.06$ mag ($M_V=-0.1$ mag, $\Delta V^{bump}_{HB}=-0.62\pm0.11$),
and that the Draco dSph shows no detectable RGB bump.
\citet{mon02} found from wide field $VI$ photometry that the Sagittarius dSph has
a single RGB bump
at $V=18.55\pm0.05$ mag ($M_V=-1.18$ mag, $\Delta V^{bump}_{HB}=+0.22\pm0.10$).
They derived a metallicity of the RGB bump, [M/H] $= -0.64\pm0.12$ dex.
\citet{bel03} reports a detection of two RGB bumps at $V\approx 21.35$ and 21.8 mag in the Leo II dSph.

In summary, among the dSphs where RGB bumps were searched for,
there are three galaxies for which two RGB bumps were suggested to have been detected
(Sculptor, Sextans and Leo II),
two galaxies for which one RGB bump was detected (Ursa Minor, Sagittarius), and
one galaxy which showed little evidence (Draco).
The brighter bump is brighter than the fainter RGB bump
by 0.6 mag for Sextans ([Fe/H]=--2.1 dex),
by 0.55 mag for Sculptor ([Fe/H]=--1.7 dex),
and by 0.45 mag for Leo II ([Fe/H]=--1.6 dex).
This indicates that the $V$ magnitude difference between the brighter bump and fainter bump
decreases with the increasing metallicity, which is contrary to the theoretical
prediction for the relation between the metallicity and
$V$ magnitude difference between the AGB bump and the RGB bump\citep{fer99}.
Further studies are needed to investigate the properties of the RGB bump and AGB bump
in dSphs.

\section{Luminosity Function of the RGB and the Main Sequence}

We have derived the $V$-band luminosity function of the RGB and the MS
stars in the Sextans dSph, which is displayed in Fig. 15.
In deriving the luminosity function we subtracted
the foreground and background contribution using the photometry of the control
field close to Pal 3 (as described in Sec. 3), and corrected the
luminosity function for the incompleteness
using the completeness data given in Table 5.

We have used
 the stars  fainter than $V=23.2$ mag with $0.2<(V-I)<1.0$
 to derive the luminosity function of the MS stars,
and the stars brighter than $V=23.2$ mag  within the $\Delta(V-I)=\pm0.1$ mag boundary of
the fiducial sequences to derive the luminosity function of the giant stars.
The logarithmic luminosity function for $V>23.2$ mag was shifted by --0.19 to match that
for $V<23.2$ mag at $V=23.2$ mag.

In Fig. 15, we plot, for comparison, the luminosity function of the metal-poor Galactic
globular cluster M92 ($5'<R<24'$) based on the $VI$ photometry given by \citet{khlee03},
who used the same CFH12K as used in this study of the Sextans dSph.
Stars brighter than $V=13.8$ mag in M92 (corresponding to $V=19.0$ mag of the Sextans dSph)
were saturated in the images used by \citet{khlee03}. 
Foreground and background contaminations for M92 were subtracted
using the control field for M92.
The luminosity function of M92 was arbitrarily shifted vertically to match that
of the Sextans dSph at $V=23.5$ mag and shifted horizontally according to the distance
of the Sextans dSph in Fig. 15.

Fig. 15 shows that
the luminosity function of the Sextans dSph is generally similar to that of M92
in several aspects:
the slope of the MS, 
the slope of the SGB and RGB,  and even the position of the RGB bump (at $V=19.95$ mag).
This result is also consistent with the fact that the fiducial sequences of
the Sextans dSph is very similar to that of M92 as shown in Fig. 11
(except for the presence of the prominent RHB in the Sextans dSph).
However, there is seen some difference as well.
Some slight excess of the Sextans stars over those of M92 is seen in the SGB and the beginning
part of the MS ($20.7<V<23.2$ mag).
This indicates that there exist some populations with younger age
in the Sextans dSph, in addition to the very old population like that in M92.

There is seen also some excess in the faint end of the MS ($V>24$ mag) of the Sextans dSph, showing
that the luminosity function of the faint MS of the Sextans is slightly steeper than that of M92.
At this faint end, the subtraction of the background objects is not perfect for the Sextans
dSph (because the data for our control field used here is only approximate),
while the luminosity function of  M92 given by \citet{khlee03} is considered to be more reliable.
Therefore the excess at the faint end may be due to imperfect subtraction of the
background objects, or they may be an intrinsic feature of the Sextans dSph. Further studies are
needed to clarify this point.

\section{Blue Stragglers}

One remarkable feature seen in the CMD of the Sextans dSph is the existence of a larger number of
blue straggler stars (BS).
Blue stars with $-0.3<(V-I)<0.4$ and $21<V<23$ mag are considered to be mostly BSs.
While there are 231 stars in this range in the CMD of the Sextans dSph, there are 8 stars
in the same range in the CMD of the control field (Fig. 6).
Therefore the contribution due to field
stars in the BS sample is only 3\%.
The BSs in the Sextans dSph were discovered early by \citet{mat95},
who covered a smaller field that the field used in this study.
We have investigated the spatial distribution and luminosity function of these BSs in the Sextans
dSph, providing important clues to understanding the origin of the BSs in the dSphs.

\subsection{Spatial distribution}

Fig. 16 displays the spatial distribution of the 238 BSs with $-0.3<(V-I)<0.4$ and
$21.1<V<23.0 $ mag.
We have divided the BSs into two groups to investigate any systematic difference:
bright BSs ($21.1<V<22.3$ mag) and faint BSs ($22.3<V<23.0$ mag).
Fig. 16 shows immediately that the bright BSs (filled circles) are more centrally concentrated
than the faint BSs (open circles), and that there is no BS within $R=100$ arcsec.

In Fig. 17 we plot the radial variation of N(BS)/N(SGB). N(BS)/N(SGB) is a ratio
of the number of the BSs and that of the SGB stars in the same magnitude
range ($21.1<V<23.0$ mag), indicating a relative variation of the BSs with respect
to the SGB stars.
Fig. 17 shows two notable features.

First, there is seen little radial variation of N(BS)/N(SGB)
for the entire sample of the BSs (black circles), telling that the radial distribution of the BSs
is similar to that of the SGB stars. This result indicates that the BSs in the Sextans dSph may be
old and of the same kind of BSs as those seen in old globular clusters.

Secondly, N(BS)/N(SGB) for the bright BSs is dramatically
different from that for the faint BSs,
N(BS)/N(SGB) for the bright BSs is steeply decreasing with increasing radius
up to at $R<1000$ arcsec (showing a strong central concentration),
and may get constant at $R>1000$ arcsec.
On the other hand, N(BS)/N(SGB) for the faint BSs is the lowest
at the innermost region ($100$ arcsec $<R<300$ arcsec)
and is approximately constant at a higher level at the outer region.

\subsection{Luminosity function}

We have derived the luminosity functions of the BSs as shown in Fig. 18.
In Fig. 18, we plotted the luminosity function of the BSs located in the inner
region ($R<11$ arcmin, blue histogram) and those in the outer region ($R>11$ arcmin, red histogram),
respectively.
The luminosity functions of these BSs keep increasing up to $V=22.9$ mag,
after which it decreases slightly up to $V=23.1$ mag.
It keeps increasing again beyond $V=23.1$ mag, where the MS stars are dominating.
It is difficult to distinguish between the BSs and MS stars for $V>23.0$ mag so that
we consider only BSs brighter than $V=23.0$ mag.

In Fig. 18, we also compared the luminosity function of the BSs in the Sextans
with the combined luminosity function of 425 BSs in 21 Galactic globular clusters
given by \citet{fus92} (see their Fig. 6).
The luminosity function of the BSs in the Sextans is found to be
roughly similar to the bright part of the luminosity function of the BSs in the Galactic globular
clusters. The Sextans dSph has a small number of BSs
a few tenths magnitude brighter than the brightest BSs in the Galactic globular clusters.

In addition, Fig. 18 shows that there is a significant difference in the luminosity functions
of the BSs between in the inner region and in the outer region:
the luminosity function of the BSs in the inner region extends to a few tenths mag
brighter magnitude (reaching $V\approx 21.2$ mag)
and has a flatter slope, compared with that of the outer region.
Linear fits to the logarithmic luminosity functions yield
a slope of $0.71\pm0.11$ for the inner region ($21.3<V<22.9$ mag)
and a slope of $1.05\pm0.20$ for the outer region ($21.7<V<22.9$ mag).
The trend is similar to the case seen in M3 which shows uniquely a bimodal distribution of BSs
\citep{fer93, bol93, bai95, fer97}.

Thus these evidences (the spatial distribution in Fig. 16, the radial distribution
in Fig. 17, and the luminosity function in Fig. 18) show consistently that the brighter BSs
in the Sextans dSph are more strongly centrally concentrated toward the center of the galaxy,
while the fainter BSs
are depleted in the central region (at $R<5$ arcmin) and are spread as the old low-mass stars like
SGB stars.

\subsection{Nature of the Blue Stragglers}

BSs have been found in open clusters, globular clusters and several dSphs
since its first discovery in M3 by \citet{san53} \citep{bai95}.
However, the origin of the BSs is not clearly known, and
the nature of the BSs in the dSphs, in particular, has been controversial.
The BSs in dSphs may be of the same kind of old BSs as these seen in the globular clusters
or they may be of different kind, i.e., normal MS stars with intermediate-age
\citep{mat95, car02}.
In this section we discuss the first possibility: the BSs in the Sextans dSph are old BSs as seen
in globular clusters, and the second possibility in the following section.

In the case of the  BSs in Galactic globular clusters,
it is generally believed today that
the BSs in dense environments like globular clusters might have formed
via mergers of primordial binaries and/or mergers of stellar collisions \citep{bai95, bai95b, sig94}.
It is expected that the BSs in the inner region of clusters are preferentially brighter and
bluer than the outer BSs, because there are generally more massive stars,
 due to dynamical mass segregation, in the inner region of the cluster
than in the outer region.
The BSs formed via collision of these massive stars in the inner region will get brighter
and hotter (due to reduced envelope opacity in the merger remnant)
than those in the outer region of the cluster \citep{bai95}.

However, we note that there are a few fundamental differences  between the Sextans dSph and
Galactic globular clusters
with regard to the origin of BSs. 
First, the central luminosity density of the Sextans dSph
(0.002 $L_\odot$ pc$^{-3}$, \citep{mat98})
is significantly lower than that of globular clusters (e.g., 3200 $L_\odot$ pc$^{-3}$ for M3, \citet{har96})
so that the collision rates
must be much lower and the dynamical evolution must be much slower in the Sextans dSph
than in globular clusters. 
Therefore it is not expected that mass segregation of the BSs results
from dynamical evolution in the central region of the Sextans dSph.
Secondly, the radial variation of the relative frequency of the BSs in galactic globular clusters
is seen over several times  the core radius ($r_c=25''$) (see Fig. 8 in \citet{fer97}),
while that of the Sextans dSph is seen within one core radius ($r_c=16.6'$).
A large core radius and the presence of a larger number of BSs in the Sextans dSph allow
us to investigate the radial variation of the BSs within the core radius in detail,
which is difficult for the case of the globular clusters.
Further studies with detailed modeling are needed to understand the observational results presented here
and investigate the origin of the BSs in the Sextans dSph.

\section{Ages of Stellar Populations}

In Fig. 19, we estimate roughly ages of the stellar populations in the Sextans dSph,
overlaying theoretical isochrones in the CMDs.
Theoretical isochrones were shifted according to the distance and reddening of the Sextans
as derived in this study.
The theoretical isochrones in Fig. 19 are Padova isochrones
based on the convective overshooting models
for the metallicity Z=0.0004 
and ages of 1, 2, 4, 7.9, 10, and 12.6 Gyr \citep{gir02}.
The metallicity $Z=0.0004$ corresponds to [Fe/H]=--1.7 dex for [$\alpha$/Fe]=0.02 as in the Sextans
dSph \citep{she01}. This value is higher than the metallicity of the Sextans, [Fe/H]=--2.1 dex,
but the isochrones with $Z=0.0004$ fits better the RGB part of the Sextans dSph
(that of M92 with [Fe/H]=--2.24 dex as well) than those with lower metallicity.
So we used the isochrones with $Z=0.0004$ in Fig. 19.

The fiducial sequences of M92 (BHB, lower RGB, and MS; blue line) 
are also plotted for comparison.
The fiducial sequences of the RGB and the MS of the Sextans dSph
agree well with those of M92,
indicating that the age of the Sextans dSph is very similar to that of M92 which is
one of the typical old halo globular clusters in our Galaxy.

Fig. 19(a) shows
that the fiducial sequences of the SGB and MSTO of the Sextans
are reasonably matched by the isochrone of 12.6 Gyr.
Some MS stars just above the fiducial sequence of the MS are fit approximately
by the isochrone of 10 Gyr.
Therefore most of the MS stars in the Sextans dSph are estimated to be formed before 10 Gyr,
mostly at about 13 Gyr.
A small number of anomalous Cepheid stars are fit roughly by the blue loop part
(pulsational instability region) of the isochrone of 1 Gyr in Fig. 19(b).

In Fig. 19(b) the position of the BS stars  
is  consistent with the MS part of the isochrones of 2 to 6 Gyr.
If these BS stars are not the same kind as the BSs in globular clusters, as suggested by \citet{mat95},
then these stars may be intermediate-age MS stars formed 2 Gyr to 6 Gyr ago.
If so, it is expected  that some red giant clump stars are present between the RHB and RGB
(but a slightly below the RHB), which are not seen clearly in the CMD.
Therefore it is concluded that the possibility that these BSs are intermediate-age normal MS stars
is very low.
\citet{car02} also concluded that the BS populations in the Ursa Minor dSph
are probably BSs originating in the old population, not the intermediate-age MS stars,
from the relative amount and spatial distribution of those stars.
It still remains to explain why there are relatively much more BSs in the dSphs
than in the galactic globular clusters.

\section{Population Gradients}

Dwarf spheroidal galaxies look monotonous apparently and seem to be composed
of a single homogeneous component. However, there are mounting evidence that dwarf galaxies
are not only more complex than expected, but also they show significant spatial variation
(or radial gradient) of the stellar populations.
Recently \citet{har01} presented a systematic and homogeneous analysis of
population gradients in several dSphs in the Local Group including the
Sextans dSph.
We have searched for any radial variation of several stellar populations in
the Sextans dSph.

First, we divide the entire CFH12K field into four different regions depending
on the radius from the center of the Sextans dSph:
Region 1 at $R<7.3$ arcmin (Central region),
Region 2 at $7.3$ arcmin $<R<11$ arcmin (Inner region),
Region 3 at $11$ arcmin $<R<15$ arcmin (Intermediate region), and
Region 4 at $R>15$ arcmin (Outer region).
The boundaries for the regions were determined so that the numbers of
the mid-RGB stars with $0.8<(V-I)<1.1$ and $18<V<21$ mag are the
same among the regions for easy comparison (and it turned out that the numbers of
the bright MS stars are also almost the same among the regions).
The areas of the central region, inner region, intermediate region and outer region
 are 167, 213, 327, and 469 arcmin$^2$, respectively.
Fig. 20 displays the \VVI diagrams of these four regions.

Secondly, we select several stellar populations of interest in the
CMD, as marked by the boxes in Fig. 20:
AGB bump, AGB group, RGB bump, BHB, RR Lyrae, RHB, bright BS
(BS1), faint BS (BS2), MS, blue RGB/red RGB and blue SGB/red SGB
(bluer and redder than the fiducial sequence of the RGB and SGB, respectively),
and a sample of foreground stars  with $1.5<(V-I)<2.5$ and $17<V<22$ mag. 
Fig. 20 plots also the fiducial sequences of the RGB and MS,
with the boundaries of $\Delta(V-I)=\pm0.1$ mag around the fiducial sequence.

We have derived the radial variation of the number of these selected
stellar populations in the four regions, which is plotted in Fig. 21 and listed in Table 9.
Note that the numbers of the mid-RGB and MS stars are almost the same among
the regions, so that the numbers of each population in Fig. 21 represent
the relative numbers with respect to the number of the mid-RGB and MS stars
in each region. The number of foreground stars is roughly proportional to the area
of each region (see Fig. 21(f)).

Figs. 20 and 21 show clear differences in several aspects among the regions:
(1) The luminosity of the bright BSs decreases as the galactocentric radius increases,
from $V=21.2$ mag in the central region to $V=22.0$ mag in the outer region, as discussed in
Sec. 10;
(2) The number of the bright BSs decreases as the radius increases, as seen in Sec. 10,
while the number of the faint BSs increases;
(3) While the SGB stars (with $20.5<V<22.5$ mag) are located mostly
in the red side of the fiducial sequence in the central region,
the SGB stars are mostly in the blue side of the fiducial sequence in the
outer region;
(4) The RGB stars (with $16.2<V<20.5$ mag) show a similar trend to that of
the SGB stars in that the RGB stars in the central region are redder than those
in the outer region, but less clearly due to the much smaller number of RGB stars;
(5) The number of the RHB stars decreases significantly as the radius increases, while
the numbers of the BHB stars and RR Lyrae variables (candidates) increase slightly;
(6) The RGB bump (bump 1) is clearly seen in the central region, while
the brighter AGB bump (bump 2) is mostly seen in the inner region and the number of  the AGB group
change little. However, due to
the small numbers of the RGB bump, AGB bump stars and AGB stars, the uncertainty of this trend
is large.

We have also investigated the population gradient using the cumulative distribution
functions and the Kolmogorov-Smirnov test.
Fig. 22 shows
the cumulative distribution functions  versus radial positions
of these selected populations.
In Fig. 22,
the dashed line represents the foreground stars which are
uniformly distributed, and the black solid line represents the MS
stars ($23.2 < V < 23.5$ mag) of the Sextans dSph
which are centrally concentrated around the center
of the galaxy.
The cumulative distribution functions of the fainter MS with ($23.6 < V < 24.0$ mag),
though not plotted,
looks almost the same as that plotted in Fig. 22.
These two lines can be used as guide lines for inspecting radial variations of other
populations.

We have performed a Kolmogorov-Smirnov test to quantify the probability
that two selected populations are drawn from the same sample (K-S probability),
the results of which are listed in Table 10.
Fig. 22 and Table 10 show several conclusions which are consistent with those
based on Figs. 20 and 21:
(1) The red RGB stars are more strongly centrally concentrated than the blue RGB stars.
The K-S probability for the red RGB and the blue RGB is zero,  $3.98\times 10^{-9}$ \%.
The former is more centrally concentrated than the MS, and the latter is more
extended than the MS;
(2) The RHB stars are more strongly centrally concentrated than the BHB stars and
the RR Lyraes.
The K-S probability for the RHB and the BHB is only 5.7 \%.
The former is more centrally concentrated than the MS, and the latter is more
extended than the MS;
(3) The RR Lyraes are similar  to the BHB stars in the radial distribution
at $R<800$ arcsec, but they are more extended out than the BHB stars at $R>800$ arcsec;
(4) The bright BSs are more strongly centrally concentrated than the faint BSs.
The K-S probability for the bright and faint BSs is close to zero, 0.4\%.
The K-S probability for the bright BSs and the RHB is 57.0 \%, and
the K-S probability for the faint BSs and the BHB is 51.7 \%. This indicates that
the bright BSs are strongly correlated with the RHB, while the faint BSs are with
the BHB.
(5) The RGB bump stars are  more strongly centrally concentrated than
the AGB bump at $R<450$ arcsec, but the latter are
more strongly centrally concentrated than the former at $R>450$ arcsec;
The K-S probability for the RGB bump  and the AGB bump is 21.3 \%.
(6) The RGB bump follows the RHB, while the AGB bump follows the BHB.
The K-S probability for the RGB bump and the RHB is 75.1\%,
and the K-S probability for the AGB bump  and the BHB is 89.0 \%.

In summary,
there are seen significant radial gradients of various kinds of stellar populations in the Sextans dSph:
the red RGB, red SGB, RHB, bright BSs, and RGB bump are, respectively,
more strongly centrally concentrated than the blue RGB, blue SGB, BHB, faint BSs and
AGB bump.
Our results for the RGB and HB are consistent with the findings of \citet{har01} who used
Washington $CT_1$ photometry.

What is the origin of the population gradient seen in the Sextans dSph?
\citet{har01} discussed in detail several possibilities to explain the population gradients seen in several
galaxies in the Local Group (including the Sextans dSph), concluding that the metallicity is a main driver
for the population gradients in some galaxies (Sextans, Sculptor, Tucana, and Andromeda VI), and
age is a critical factor in Carina.
In forthcoming paper, we are planning to investigate in detail
the origin of this population gradient using the star formation history models
(Lee et al. 2003, in preparation).

\section{Discussion}

\subsection{Do Multiple Old Stellar Populations Exist in the Sextans dSph?}

In the study of the Sculptor dSph, an analog to the Sextans dSph, \citet{maj99} suggested,
that the metallicity distribution of the stars in the Sculptor dSph is bimodal
(one at [Fe/H]$\sim -2.3$ dex and another at [Fe/H]$\sim -1.5$ dex),
based on a discontinuity in the $V$-band luminosities of the HB stars
(the blue edge of the RHB is about 0.15 mag fainter than the red edge of the BHB) and
the presence of two RGB bumps
(a blue bump at $V_0=19.3$ mag and a red fainter bump at $V_0=20.0$ mag)
(see their Fig. 3).
The metal poor populations (BHB, blue RGB, blue bright RGB bump) are spatially more
extended than the metal rich populations (RHB, red RGB, red faint RGB bump), and the HB
index increases by 0.4 from the center to the outer region at $R\sim500$ arcsec.
In addition, they suggested that the metal-poor population formed in an earlier, more extended
burst.%

Motivated by the \citet{maj99}'s study on the Sculptor dSph, \citet{bel01a} suggested,
from wide-field ($33\times 34$ arcmin$^2$) $BVI$ CCD photometry (reaching to $I=20.5$ mag),
that the Sextans dSph also has two old populations: a major one with [Fe/H]$=-1.8$ dex
and a minor one with [Fe/H]$< -2.3$ dex. This conclusion is based on
three points: (1) there is a BHB which appears to lie on a
brighter sequence than the RHB and the RR Lyrae stars in their $I-(B-I)$ diagram;
(2) there are hints that the $(B-I)$ color distribution of the RGB stars
with $16<I<20.5$ mag is  bimodal, with a color difference of $\Delta (B-I)= 0.04$; and
(3) two RGB bumps are found to be at $B=20.2$ and 20.8 mag.

However, our results  are not consistent
with the above points given by \cite{bel01a} as follows.
First,
the red edge of the BHB in the Sextans dSph is not brighter than,
but is rather slightly fainter
than the blue edge of the RHB in the $I$-magnitude (see Fig. 11),
and the former is very similar to the latter in the $V$ magnitude
(the difference is only $\Delta V=0.03$ mag) (see Fig. 10 and Sec. 6).
This result is the opposite to the first point of \citet{bel01a}.
A small number of stars seen slightly above the red edge of the BHB in Fig. 3 turned out to be
mostly RR Lyrae variables (as shown in Fig. 8).
Note the contrast between the CMDs with variable stars (Fig. 3 and 8) and without variable stars (Fig. 10).
Thus the first point of \citet{bel01a} is not supported by our result.
In addition, this shows that there is a difference between the Sextans and the Sculptor
in the luminosity difference between the BHB and the RHB.

Secondly, the $(V-I)$ color distribution of the RGB (excluding the AGB) in Fig. 13 shows
little hint of bimodality for $I>18$ mag
and that the color distribution is roughly constant over the range of bright magnitude
($15.9<I<18$ mag)
(but the number of stars is too small to do reliable Gaussian fitting).
On the other hand,
\citet{bel01a} suggested that the $(B-I)$ color distribution is bimodal (though weakly) over
the range  $16<I<20.5$ mag,
and that the color differences of the two peaks are the same ($\Delta (B-I)= 0.04$)
for three ranges of magnitudes,
$16.0<I<18.5$,  $18.5<I<19.5$, and $19.5<I<20.5$ mag (see their Fig. 3).
If these two weak peaks are due to two populations with [Fe/H]=--1.8 and $<-2.3$ dex,
it is expected that the color difference of the two peaks should decrease
as the magnitude increases
(e.g., see Fig. 10 and the fiducial sequences of Galactic globular clusters in Fig. 11).
In addition, the metallicity distribution of 43 RGB stars in the Sextans dSph
based on the spectroscopic observations by \citet{sun93}
does not show any hint of bimodality,
although it shows a large range of [Fe/H] from $-2.52$ to --1.57 dex  (see their Fig. 9).
Therefore the bimodal color distribution of the RGB suggested by \citet{bel01a},
if it exists, does not necessarily lead to the existence of two old populations
with [Fe/H] = --1.8 and $ <-2.3$ dex.
Finally, the brighter bump  seen on the blue RGB is probably an AGB bump,
not an RGB bump, as discussed in Sec. 8.

In summary, our results support none of the three points \citet{bel01a}
used for suggesting that the Sextans dSph also has two old populations.
However, a large spread in the color distribution of the upper RGB shows
that there is a metallicity dispersion of $\Delta$[Fe/H]$\approx 0.2$ dex, indicating
that the major star formation must have continued over a few Gyr with increasing metallicity.

\subsection{Is the Sextans Peculiar among the Local Group dSphs?}

The Sextans dSph has been considered as a distinguishable outlier among the dSphs in the
Local Group because of its metallicity, integrated luminosity and the horizontal branch index
(starting from the earliest studies of \citet{mat91} and \citet{dac91}).
The HB index which is defined as (N(BHB)-N(RHB))/(N(BHB)+N(RR)+N(RHB)) is
a very useful parameter to represent the characteristics of
HB populations in globular clusters and resolved galaxies \citep{lee90}.
We have derived the value of the HB index of the Sextans
dSph, as listed in Table 9.
We discuss this point using the most recent data of the Local Group dwarf galaxies.

Fig. 23(a) shows the [Fe/H] versus $M_V$ diagram of the Sextans
in comparison with other Local Group dwarf galaxies (symbols) and Galactic globular clusters (dots).
The morphological types of the dwarf galaxies are dSphs (filled squares),
dwarf irregulars (dIrrs, pentagons), dwarf ellipticals (dEs, filled circles), and
transition types between dSph and dIrr (open triangles).
The data used for Fig. 23 are based on the compiled data of \citet{mat98}, \citet{har01}
and \citet{gre03},
including our estimates of [Fe/H] and the HB index for the Sextans.

It has been known long that the mean metallicity of the dSphs increases with the increasing
luminosity, in contrast to the Galactic globular clusters, as shown in Fig. 23(a).
The Sextans dSph has the lowest metallicity among the dSphs.
A linear least squares fit to the data of metallicity and the luminosity for the early type galaxies
(the dSphs and dEs)
yields [Fe/H]$=-0.16(\pm0.02)M_V -3.39(\pm0.19)$ with rms of 0.17.
The Sexans has a deviation of $\Delta$[Fe/H]$=-0.23$ dex, only slightly larger than one sigma.
Thus the Sextans dSph follows reasonably well the [Fe/H]--$M_V$ relation
of other dSphs and dEs including the Sculptor.

Fig. 23(b) displays that [Fe/H] versus the HB index diagram for the Sextans dSph in comparison with
other Local Group dwarf galaxies (symbols) and Galactic globular clusters (dots).
Theoretical isochrones based on the HB models for relative ages of --1.1, 0.0 and 1.1 Gyr
\citep{rey01} are also overlayed.
The range of the HB index and the metallicity dispersion of the Sextans are shown by the
horizontal and vertical lines centered on the position of the Sextans. 
The Sextans shows the largest deviation (toward the lowest metallicity) from the main group,
while the Sculptor is located within the main group as well as most Galactic globular clusters.
A metallicity increase by 0.3--0.4 dex is needed to put the Sextans within the main group.
If we extrapolate the isochrone to the lower metallicity end, the relative age of the Sextans
will be about --2 Gyr, showing that it is about two Gyr younger than the Sculptor.
However, this age difference is not consistent with the result from the main sequence age estimates
(i.e. the MSTO age of the Sextans is similar to that of M92).
In this sense, the Sextans dSph is a peculiar dSph in the Local Group and it remains to be
explained.

\subsection{A Brief History of Star Formation of the Sextans dSph}

Here we present a schematic description of the star formation history of the Sextans dSph
using observational clues which are summarized in the final section.
The majority of stars in the Sextans were formed at the similar time to the
age of metal-poor globular clusters like M92, and they are very metal-poor
([Fe/H] = --2.1 dex). 
The metallicity spread ($\Delta$[Fe/H]$\approx 0.2$ dex) observed in the
RGB, and the significant population gradients indicate
that this star formation might have continued for a few Gyr, which is
also consistent with the presence of strong RHB and the presence of bright MS stars
slightly above the MSTO of the main population (see also \citet{iku02}).
A small amount of stars, including anomalous Cepheids, were formed
later about one Gyr ago.

\section{Summary}

We present deep wide field  $VI$ CCD photometry of the Sextans dwarf spheroidal galaxy
in the Local Group, covering a field of $42' \times 28'$ located at the center of
the galaxy (supplemented by short $B$ photometry). These data allowed us to investigate
in detail the properties of several stellar populations located in the wide region
of the Sextans dSph.
Primary results are summarized below.

\begin{enumerate}

\item Color-magnitude diagrams of the Sextans dSph show well-defined RGB, BHB,
prominent RHB, AGB,
about 120 variable stars including RR Lyraes and anomalous Cepheids, about 230 BSs,
and MS stars.

\item The MSTO of the old population is found to be located at $V\approx 23.7$ mag and $(V-I)\approx 0.56$.

\item The distance to the galaxy is derived using the $I$-band magnitude of the
tip of the RGB at $I$(TRGB)$=15.95\pm0.04$:
$(m-M)_0=19.90\pm0.06$ for an adopted reddening of $E(B-V)=0.01$.
This estimate agrees well with the distance estimate based on the mean $V$-band magnitude
of the HB at $V$(HB)$=20.37\pm0.04$

\item The mean metallicity of the RGB is estimated from the $(V-I)$ color at $M_I=-3.5$ and --3.0 mag:
[Fe/H]$=-2.1\pm0.1$(stat. error)$\pm0.2$ (standard calibration error) dex,
with  a dispersion of $\sigma$[Fe/H]=0.2 dex.

\item There is found to be one RGB bump at $V=19.95$ mag ($M_V=0.03$ mag),
and a weak brighter bump at $V=19.35$ mag($M_V=-0.58$ mag) which is probably an AGB bump.
The $V$ magnitude differences between the bumps and the HB are $\Delta V^{bump}_{HB}=-0.45$
for the RGB bump, and --1.0 mag for the brighter AGB bump.

\item The $V$-band luminosity function of the BSs in the inner region
is found to extend to a brighter magnitude  and to have a slightly flatter slope
compared with that of the BSs in the outer region.
The bright BSs are more strongly centrally concentrated than the faint BSs.
The faint BSs are depleted in the central region
and the spatial distribution of the faint BSs
is similar to that of the SGB stars.
The BSs in the Sextans are probably old BSs as seen in Galactic globular clusters, rather than the
intermediate-age MS stars.

\item The $V$-band luminosity function of the RGB and MS stars in the Sextans dSph
is in general similar to that of
the globular cluster M92, with a slight excess of stars
in the magnitude range brighter than the MSTO with respect to that of M92.

\item The age of the MSTO of the main population is estimated to be similar to that of
the metal-poor Galactic globular clusters like M92, and there are seen some stellar populations
with younger age.

\item Significant radial gradients are seen for several populations:
the RHB, the red RGB, the red SGB, and the bright BSs are more
centrally concentrated toward the center of the galaxy, compared with
the BHB, the blue RGB, the blue SGB, and the faint BS, respectively.

\end{enumerate}

\acknowledgments
The authors thank anonymous referee for improving this paper.
M.G.L thanks the staff of
the Department of Terrestrial Magnetism, Carnegie Institution of Washington,
for their kind hospitality.
M.G.L. was supported in part by the Korean Research Foundation Grant
(KRF-2000-DP0450).
This work is partially supported by the BK21 program.
Y. W. L., Y. J. S., and S. C. R. acknowledge the support for this work provided
    by the Creative Research Initiative Program of the Korean Ministry of Science and Technology.
The authors are grateful to the staff members of the CFHT and the Bohyunsan
Optical Astronomy Observatory.

\clearpage


\begin{deluxetable}{ccccccl}
\tablecaption{Observation Log. \label{tbl-1}}  
\tablewidth{0pc}
\tablehead{
\colhead{Field} & \colhead{Filter} & \colhead{T(exp)} & \colhead{Airmass} & \colhead{Seeing} &
\colhead{Date(UT)} }
\startdata
CFH12K-F & $V$ & $3 \times 1200$ sec & 1.51 & 1\farcs3 & Feb 16, 2001 \\
CFH12K-F & $I$ & $3 \times 600$ sec & 1.25 & 1\farcs1 & Feb 16, 2001 \\
CFH12K-S & $B$ & 60 sec & 1.29 & 1\farcs1 & Feb 17, 2001 \\
CFH12K-S & $V$ & 60 sec & 1.32 & 1\farcs0 & Feb 17, 2001 \\
CFH12K-S & $I$ & 60 sec & 1.30 & 0\farcs8 & Feb 17, 2001 \\
BOAO-F2 & $B$ & 300 sec & 1.34 & 1\farcs5 & Nov 19, 2001 \\
BOAO-F2 & $V$ & 150 sec & 1.33 & 1\farcs4 & Nov 19, 2001 \\
BOAO-F2 & $I$ & 75 sec & 1.35 & 1\farcs0 & Nov 19, 2001 \\
BOAO-F1 & $B$ & 300 sec & 1.42 & 2\farcs4 & Nov 20, 2001 \\
BOAO-F1 & $V$ & 150 sec & 1.40 & 1\farcs9 & Nov 20, 2001 \\
BOAO-F1 & $I$ & 75 sec & 1.38 & 1\farcs9 & Nov 20, 2001 \\
BOAO-F3 & $B$ & 300 sec & 1.36 & 2\farcs4 & Nov 20, 2001 \\
BOAO-F3 & $V$ & 150 sec & 1.34 & 1\farcs9 & Nov 20, 2001 \\
BOAO-F3 & $I$ & 75 sec & 1.33 & 1\farcs9 & Nov 20, 2001 \\
BOAO-F4 & $B$ & 300 sec & 1.26 & 2\farcs2 & Dec 30, 2002 \\
BOAO-F4 & $V$ & 150 sec & 1.27 & 2\farcs1 & Dec 30, 2002 \\
BOAO-F4 & $I$ & 75 sec & 1.26 & 2\farcs1 & Dec 30, 2002 \\
BOAO-F5 & $B$ & 300 sec & 1.27 & 2\farcs2 & Dec 30, 2002 \\
BOAO-F5 & $V$ & 150 sec & 1.28 & 2\farcs2 & Dec 30, 2002 \\
BOAO-F5 & $I$ & 75 sec & 1.27 &  2\farcs1 & Dec 30, 2002 \\
\enddata
\end{deluxetable}


\begin{deluxetable}{ccccc}
\tablecaption{Zero points of CFH12K$^{\rm a}$. \label{tbl-2}}  
\tablewidth{0pc}
\tablehead{
\colhead{Chip} & \colhead{Zero($B$)} & \colhead{Zero($V_{bv}$)} & \colhead{Zero($V_{vi}$)} & \colhead{Zero($I$)} }
\startdata
   0  &  0.187  &  0.080 &  0.081  &  0.082 \\
   1  &  0.088  &  0.015 &  0.013  &  0.050 \\
   2  &  0.031  & -0.037 & -0.037  & -0.006 \\
   3  &  0.023  & -0.013 & -0.013  &  0.054 \\
   4  &  0.077  &  0.025 &  0.026  &  0.080 \\
   5  &  0.126  &  0.055 &  0.060  &  0.096 \\
   6  &  0.112  &  0.062 &  0.063  &  0.119 \\
   7  &  0.009  & -0.011 & -0.009  &  0.020 \\
   8  & -0.012  & -0.031 & -0.030  & -0.019 \\
   9  &  0.000  &  0.000 &  0.000  &  0.000 \\
  10  &  0.047  &  0.001 &  0.003  &  0.021 \\
  11  &  0.163  &  0.097 &  0.085  &  0.071 \\
\enddata
\tablenotetext{a}{Zero points are normalized with respect to that of Chip 9.}
\end{deluxetable}



\begin{deluxetable}{ccccccccc}
\tablecaption{$BVI$ Photometry of the Sextans Dwarf Spheroidal Galaxy (electronic table). \label{tbl-3}}
\tabletypesize{\scriptsize}
\tablewidth{0pc}
\tablehead { \colhead{ID} & \colhead{RA(2000)} & \colhead{Dec(2000)}
& \colhead{$V$} & \colhead{$\sigma(V)$} & \colhead{$(V-I)$} & \colhead{$\sigma(V-I)$}
& \colhead{$(B-V)$} & \colhead{$\sigma(B-V)$}}
\startdata
    1 & 10:12:42.20 &  -1:47:41.5 &  14.77 &  0.01 &   &    &  0.62 &  0.01 \\
    2 & 10:12:44.63 &  -1:43:34.6 &  14.84 &  0.01 &   &    &  0.91 &  0.01 \\
    3 & 10:14:05.72 &  -1:34:09.8 &  14.85 &  0.01 &   &    &  0.85 &  0.01 \\
    4 & 10:11:48.52 &  -1:32:06.0 &  14.89 &  0.01 &   &    &  1.23 &  0.01 \\
    5 & 10:14:18.42 &  -1:42:11.7 &  14.89 &  0.01 &   &    &  0.64 &  0.01\\
    6 & 10:12:47.29 &  -1:38:20.2 &  14.90 &  0.01 &   &    &  0.74 &  0.01 \\
    7 & 10:12:57.51 &  -1:40:24.8 &  14.96 &  0.01 &   &    &  1.61 &  0.01 \\
    8 & 10:13:30.55 &  -1:48:52.1 &  14.96 &  0.01 &   &    &  0.03 &  0.01 \\
   10 & 10:13:41.05 &  -1:34:55.3 &  15.03 &  0.01 &   &    &  0.56 &  0.01 \\ \cutinhead{a full table will be provided electronically only.}
\enddata
\end{deluxetable}


\begin{deluxetable}{ccccc}
\tablecaption{ Mean Photometric Errors. \label{tbl-4}}
\tablewidth{0pc} 
\tablehead {
\colhead{$V$} & \colhead{$\sigma(V)$} & \colhead{$\sigma(V-I)$} & \colhead{$\sigma(B-V)$}
}
\startdata
  17.25 &  0.003 &  0.003 &  0.005  \\
  17.75 &  0.003 &  0.003 &  0.006  \\
  18.25 &  0.003 &  0.003 &  0.008  \\
  18.75 &  0.004 &  0.005 &  0.010  \\
  19.25 &  0.004 &  0.003 &  0.013  \\
  19.75 &  0.003 &  0.004 &   0.019 \\
  20.25 &  0.008 &  0.006 &   0.024 \\ 
  20.75 &  0.008 &  0.008 &   0.037 \\ 
  21.25 &  0.007 &  0.011 &   0.054 \\ 
  21.75 &  0.010 &  0.017 &   0.079 \\
  22.25 &  0.016 &  0.029 &   0.104 \\ 
  22.75 &  0.026 &  0.048 &   0.149 \\
  23.25 &  0.039 &  0.079 &   0.222 \\
  23.75 &  0.059 &  0.120 &   0.338 \\
  24.25 &  0.093 &  0.178 &   0.590 \\
  24.75 &  0.137 &  0.241 &   0.274 \\
  25.25 &  0.200 &  0.296 &   0.223 \\ 
\enddata
\end{deluxetable}


\begin{deluxetable}{cccc}
\tablecaption{Completeness of our Photometry. \label{tbl-5}}
\tablewidth{0pc} 
\tablehead{\colhead{$V$} & \colhead{$\Lambda$\tablenotemark{a}} &  \colhead{$I$} & \colhead{$\Lambda$} }
\startdata
  20.75 &   1.000 &   19.75 &   1.000 \\
  21.25 &   0.991 &   20.25 &   0.995 \\
  21.75 &   0.982 &   20.75 &   0.984 \\
  22.25 &   0.981 &   21.25 &   0.951 \\
  22.75 &   0.900 &   21.75 &   0.963 \\
  23.25 &   0.900 &   22.25 &   0.915 \\
  23.75 &   0.679 &   22.75 &   0.804 \\
  24.25 &   0.598 &   23.25 &   0.691 \\
  24.75 &   0.308 &   23.75 &   0.410 \\
  25.25 &   0.034 &   24.25 &   0.060 \\
\enddata
\tablenotetext{a}{$\Lambda =$ N(recovered)/N(added).}
\end{deluxetable}


\begin{deluxetable}{ccc ccccccl}
\tablecaption{A List of Variable Star Candidates in the Sextans Dwarf Spheroidal
Galaxy\tablenotemark{a}. \label{tbl-6}}
\tabletypesize{\scriptsize}
\tablewidth{0pc}
\tablehead { \colhead{ID} & \colhead{RA(2000)} & \colhead{Dec(2000)}
& \colhead{$V$} & \colhead{$\sigma(V)$} & \colhead{$(V-I)$} & \colhead{$\sigma(V-I)$}
& \colhead{$(B-V)$} & \colhead{$\sigma(B-V)$} & Remarks}
\startdata
VI01 & 10:11:59.65 &  -1:33:42.4 &  20.51 &  0.02 &   0.73 &  0.02 &   0.55 &  0.03 &   \\
VI02 & 10:11:46.61 &  -1:29:20.0 &  20.74 &  0.02 &   0.43 &  0.04 &   0.37 &  0.03 &   \\
VI03 & 10:11:56.39 &  -1:31:28.0 &  20.33 &  0.01 &   0.39 &  0.03 &   0.38 &  0.02 &   \\
VI04 & 10:11:44.92 &  -1:26:41.8 &  20.14 &  0.01 &   0.31 &  0.02 &   0.37 &  0.02 &   \\
VI05 & 10:11:51.16 &  -1:34:30.2 &  20.33 &  0.01 &   0.56 &  0.02 &   0.51 &  0.03 &   \\
VI06 & 10:11:42.28 &  -1:31:41.3 &  21.94 &  0.05 &   1.07 &  0.07 &        &      &   \\
VI07 & 10:12:26.56 &  -1:36:43.5 &  20.72 &  0.02 &   0.66 &  0.03 &   0.38 &  0.03 &   \\
VI08 & 10:12:29.49 &  -1:33:36.9 &  20.70 &  0.02 &   0.64 &  0.03 &   0.39 &  0.03 &   \\
VI09 & 10:12:13.08 &  -1:28:55.0 &  20.33 &  0.02 &   0.39 &  0.03 &   0.33 &  0.02 &   \\
VI10 & 10:12:29.71 &  -1:30:46.5 &  20.60 &  0.02 &   0.54 &  0.03 &   0.50 &  0.03 &   \\
VI11 & 10:12:21.61 &  -1:29:27.8 &  20.42 &  0.02 &   0.61 &  0.03 &   0.59 &  0.03 &   \\
VI12 & 10:12:31.88 &  -1:36:04.4 &  20.62 &  0.02 &   0.58 &  0.03 &   0.46 &  0.03 &   \\
VI13 & 10:12:32.29 &  -1:34:08.9 &  20.31 &  0.02 &   0.59 &  0.02 &   0.48 &  0.03 &   \\
VI14 & 10:12:17.91 &  -1:25:33.3 &  20.20 &  0.02 &   0.42 &  0.02 &   0.46 &  0.02 &   \\
VI15 & 10:12:31.23 &  -1:35:41.7 &  20.35 &  0.02 &   0.33 &  0.03 &   0.25 &  0.02 &   \\
VI16 & 10:12:16.61 &  -1:26:49.1 &  20.24 &  0.02 &   0.43 &  0.03 &   0.48 &  0.02 &   \\
VI17 & 10:12:28.74 &  -1:26:59.0 &  20.39 &  0.02 &   0.59 &  0.03 &   0.47 &  0.03 &   \\
VI18 & 10:12:12.51 &  -1:36:40.8 &  21.96 &  0.06 &   1.12 &  0.08 &        &      &   \\
VI19 & 10:13:02.50 &  -1:28:42.9 &  20.70 &  0.02 &   0.68 &  0.03 &   0.51 &  0.04 &   \\
VI20 & 10:12:52.07 &  -1:26:08.3 &  20.51 &  0.02 &   0.57 &  0.03 &   0.48 &  0.03 &   \\
VI21 & 10:12:40.37 &  -1:29:46.9 &  20.61 &  0.02 &   0.57 &  0.03 &   0.38 &  0.03 &   \\
VI22 & 10:12:53.34 &  -1:32:11.6 &  20.05 &  0.01 &   0.35 &  0.02 &   0.32 &  0.02 &   17    RRab \\
VI23 & 10:12:54.09 &  -1:33:26.0 &  20.55 &  0.02 &   0.57 &  0.03 &   0.48 &  0.03 &   18    RRab \\
VI24 & 10:12:40.97 &  -1:30:46.1 &  19.97 &  0.01 &   0.39 &  0.02 &   0.37 &  0.02 &   \\
VI25 & 10:13:29.92 &  -1:28:48.5 &  20.31 &  0.02 &   0.68 &  0.02 &   0.47 &  0.03 &   \\
VI26 & 10:13:21.47 &  -1:28:29.2 &  20.56 &  0.02 &   0.53 &  0.03 &   0.50 &  0.03 &   \\
VI27 & 10:13:31.02 &  -1:37:02.1 &  20.54 &  0.02 &   0.59 &  0.03 &   0.41 &  0.03 &    4    RRab \\
VI28 & 10:13:32.20 &  -1:23:37.3 &  20.53 &  0.02 &   0.47 &  0.03 &   0.39 &  0.03 &   \\
VI29 & 10:13:19.88 &  -1:35:56.1 &  20.62 &  0.02 &   0.54 &  0.03 &   0.43 &  0.03 &   15    RRab \\
VI30 & 10:13:32.16 &  -1:24:55.7 &  20.56 &  0.02 &   0.47 &  0.03 &   0.46 &  0.03 &   \\
VI31 & 10:13:05.87 &  -1:35:07.0 &  20.61 &  0.02 &   0.61 &  0.03 &   0.55 &  0.03 &    7    RRab \\ 
VI32 & 10:13:08.33 &  -1:34:13.0 &  20.69 &  0.02 &   0.57 &  0.03 &   0.45 &  0.03 &   10    RRab \\
VI33 & 10:13:08.25 &  -1:33:58.4 &  19.86 &  0.01 &   0.37 &  0.02 &   0.24 &  0.03 &    9     ACH \\ 
VI34 & 10:13:15.74 &  -1:34:59.4 &  20.21 &  0.01 &   0.49 &  0.02 &   0.37 &  0.02 &   14    RRab \\
VI35 & 10:13:32.46 &  -1:23:48.1 &  20.22 &  0.01 &   0.43 &  0.02 &   0.47 &  0.03 &   \\ 
VI36 & 10:13:26.74 &  -1:26:54.7 &  20.47 &  0.02 &   0.49 &  0.03 &   0.45 &  0.03 &   \\
VI37 & 10:13:58.16 &  -1:35:50.8 &  19.75 &  0.01 &   0.35 &  0.02 &   0.17 &  0.02 &    3    RRab \\
VI38 & 10:13:42.49 &  -1:27:04.5 &  19.63 &  0.01 &   0.20 &  0.02 &   0.14 &  0.01 &   \\
VI39 & 10:13:38.61 &  -1:28:02.0 &  20.14 &  0.01 &   0.50 &  0.02 &   0.37 &  0.02 &   \\
VI40 & 10:14:22.88 &  -1:31:00.8 &  20.55 &  0.02 &   0.53 &  0.03 &   0.39 &  0.03 &   \\
VI41 & 10:14:25.70 &  -1:23:32.1 &  19.92 &  0.01 &   0.41 &  0.02 &   0.39 &  0.02 &   \\
VI42 & 10:14:12.15 &  -1:31:24.9 &  20.37 &  0.01 &   0.50 &  0.02 &   0.37 &  0.02 &   \\ 
VI43 & 10:11:55.43 &  -1:46:10.1 &  21.52 &  0.04 &   0.97 &  0.05 &   0.92 &  0.08 &   \\
VI44 & 10:12:19.12 &  -1:40:24.7 &  20.63 &  0.02 &   0.65 &  0.03 &   0.51 &  0.03 &   \\ 
VI45 & 10:12:20.49 &  -1:37:33.0 &  20.35 &  0.02 &   0.57 &  0.03 &   0.46 &  0.03 &   \\
VI46 & 10:12:15.24 &  -1:37:10.2 &  20.59 &  0.02 &   0.60 &  0.03 &   0.55 &  0.03 &   \\
VI47 & 10:12:13.86 &  -1:40:48.2 &  20.78 &  0.02 &   0.62 &  0.03 &   0.51 &  0.04 &   \\
VI48 & 10:12:27.53 &  -1:47:11.1 &  20.00 &  0.01 &   0.40 &  0.02 &   0.28 &  0.02 &   \\
VI49 & 10:12:25.29 &  -1:43:18.0 &  20.20 &  0.01 &   0.46 &  0.02 &   0.38 &  0.02 &   \\
VI50 & 10:12:23.99 &  -1:41:55.9 &  20.34 &  0.02 &   0.51 &  0.03 &   0.34 &  0.02 &   \\
VI51 & 10:12:38.93 &  -1:39:04.0 &  20.68 &  0.02 &   0.62 &  0.03 &   0.46 &  0.03 &   \\
VI52 & 10:13:02.39 &  -1:44:13.8 &  20.14 &  0.01 &   0.38 &  0.02 &   0.24 &  0.02 &   29     RRc \\
VI53 & 10:12:39.07 &  -1:46:54.8 &  19.89 &  0.01 &   0.29 &  0.02 &   0.21 &  0.02 &   \\ 
VI54 & 10:12:37.68 &  -1:49:14.3 &  20.14 &  0.01 &   0.41 &  0.02 &   0.25 &  0.02 &   \\
VI55 & 10:12:38.21 &  -1:48:43.6 &  20.62 &  0.02 &   0.53 &  0.03 &   0.36 &  0.03 &   \\ 
VI56 & 10:13:01.62 &  -1:39:04.9 &  20.24 &  0.01 &   0.50 &  0.02 &   0.42 &  0.02 &   40    RRab \\
VI57 & 10:12:47.55 &  -1:39:51.9 &  20.49 &  0.02 &   0.60 &  0.03 &   0.45 &  0.03 &   \\ 
VI58 & 10:12:52.75 &  -1:39:57.0 &  21.08 &  0.03 &   0.77 &  0.04 &   0.42 &  0.04 &   \\
VI59 & 10:12:49.62 &  -1:42:12.1 &  21.57 &  0.04 &   0.75 &  0.07 &   0.64 &  0.07 &   \\ 
VI60 & 10:12:49.62 &  -1:42:12.1 &  21.57 &  0.04 &   0.75 &  0.07 &   0.64 &  0.07 &   \\
VI61 & 10:13:01.60 &  -1:42:53.6 &  22.02 &  0.07 &   0.70 &  0.10 &   0.55 &  0.10 &   \\ 
VI62 & 10:12:42.67 &  -1:48:36.8 &  22.19 &  0.07 &   0.91 &  0.10 &        &      &   \\
VI63 & 10:13:21.10 &  -1:49:09.1 &  20.14 &  0.01 &   0.30 &  0.02 &   0.14 &  0.02 &   \\ 
VI64 & 10:13:20.61 &  -1:42:44.3 &  20.66 &  0.02 &   0.63 &  0.03 &   0.43 &  0.03 &   \\
VI65 & 10:13:25.91 &  -1:50:17.7 &  20.40 &  0.02 &   0.53 &  0.03 &   0.37 &  0.02 &   \\ 
VI66 & 10:13:09.15 &  -1:37:30.4 &  20.61 &  0.02 &   0.65 &  0.03 &   0.47 &  0.03 &   11    RRab \\
VI67 & 10:13:27.25 &  -1:46:51.2 &  20.75 &  0.02 &   0.73 &  0.03 &   0.50 &  0.03 &   39    RRab \\ 
VI68 & 10:13:11.62 &  -1:42:12.8 &  19.98 &  0.01 &   0.40 &  0.02 &   0.22 &  0.02 &   30    RRab \\
VI69 & 10:13:27.50 &  -1:43:44.1 &  20.25 &  0.01 &   0.57 &  0.02 &   0.38 &  0.02 &   \\ 
VI70 & 10:13:07.02 &  -1:37:25.1 &  20.27 &  0.01 &   0.52 &  0.02 &   0.41 &  0.02 &    8    RRab \\
VI71 & 10:13:22.14 &  -1:42:55.8 &  20.40 &  0.02 &   0.59 &  0.02 &   0.42 &  0.02 &   32    RRab \\ 
VI72 & 10:13:20.42 &  -1:40:45.0 &  20.55 &  0.02 &   1.34 &  0.02 &   1.51 &  0.05 &   \\
VI73 & 10:13:36.47 &  -1:37:53.9 &  19.14 &  0.01 &   0.46 &  0.01 &   0.38 &  0.01 &    1     ACH \\ 
VI74 & 10:13:34.52 &  -1:38:42.6 &  19.37 &  0.01 &   0.51 &  0.01 &   0.37 &  0.01 &    5     ACF \\
VI75 & 10:13:43.68 &  -1:48:45.5 &  20.55 &  0.02 &   0.58 &  0.03 &   0.46 &  0.03 &   25    RRab \\ 
VI76 & 10:13:33.23 &  -1:47:57.5 &  20.52 &  0.02 &   0.45 &  0.03 &   0.29 &  0.03 &   27     RRc \\
VI77 & 10:13:49.22 &  -1:50:43.7 &  20.19 &  0.01 &   0.41 &  0.02 &   0.29 &  0.02 &   \\ 
VI78 & 10:13:36.94 &  -1:49:47.9 &  20.26 &  0.01 &   0.53 &  0.02 &   0.41 &  0.02 &   \\
VI79 & 10:13:40.17 &  -1:41:00.7 &  20.19 &  0.01 &   0.42 &  0.02 &   0.35 &  0.02 &   23 faint RR \\
VI80 & 10:13:45.05 &  -1:43:05.0 &  20.25 &  0.01 &   0.49 &  0.02 &   0.42 &  0.02 &   26    RRab \\
VI81 & 10:14:13.04 &  -1:47:09.0 &  20.54 &  0.02 &   0.60 &  0.03 &   0.59 &  0.03 &   \\ 
VI82 & 10:14:10.86 &  -1:45:36.6 &  19.78 &  0.01 &   0.33 &  0.02 &   0.28 &  0.02 &   \\
VI83 & 10:14:25.96 &  -1:41:58.9 &  20.09 &  0.01 &   0.36 &  0.02 &   0.37 &  0.02 &   \\ 
VI84 & 10:14:26.19 &  -1:40:47.4 &  20.27 &  0.02 &   0.53 &  0.02 &   0.42 &  0.02 &   \\
VI85 & 10:14:02.66 &  -1:38:46.6 &  20.58 &  0.02 &   0.55 &  0.03 &   0.50 &  0.03 &   44    RRab \\ 
VV01 & 10:11:48.05 &  -1:35:52.5 &  18.33 &  0.01 &   0.78 &  0.01 &   0.71 &  0.01 &   \\
VV02 & 10:12:27.05 &  -1:30:21.4 &  18.47 &  0.01 &   1.12 &  0.01 &   1.05 &  0.01 &   \\ 
VV03 & 10:12:15.83 &  -1:27:24.7 &  18.34 &  0.01 &   1.07 &  0.01 &   1.19 &  0.01 &   \\
VV04 & 10:12:31.02 &  -1:36:49.9 &  20.36 &  0.02 &   0.51 &  0.03 &   0.46 &  0.03 &   \\ 
VV05 & 10:12:08.46 &  -1:28:45.5 &  20.33 &  0.02 &   0.38 &  0.03 &   0.39 &  0.02 &   \\
VV06 & 10:12:32.15 &  -1:26:48.4 &  22.76 &  0.13 &   1.69 &  0.14 &        &      &   \\
VV07 & 10:12:49.57 &  -1:35:11.7 &  18.46 &  0.01 &   1.29 &  0.01 &   1.11 &  0.01 &   \\
VV08 & 10:12:59.37 &  -1:27:03.1 &  20.13 &  0.01 &   0.43 &  0.02 &   0.43 &  0.02 &   \\
VV09 & 10:13:08.74 &  -1:23:11.3 &  19.95 &  0.01 &   0.63 &  0.02 &   1.07 &  0.03 &   \\
VV10 & 10:13:27.74 &  -1:32:43.7 &  20.52 &  0.02 &   0.48 &  0.03 &   0.30 &  0.03 &   16 faint RR \\
VV11 & 10:13:15.72 &  -1:32:43.5 &  22.27 &  0.08 &   0.22 &  0.14 &        &      &   \\
VV12 & 10:13:30.26 &  -1:29:35.0 &  22.19 &  0.07 &   0.92 &  0.09 &   0.62 &  0.13 &   \\ 
VV13 & 10:13:47.41 &  -1:36:19.7 &  20.33 &  0.02 &   0.31 &  0.03 &   0.22 &  0.03 &   \\
VV14 & 10:13:39.51 &  -1:26:33.3 &  20.38 &  0.02 &   0.48 &  0.03 &   0.31 &  0.03 &   \\
VV15 & 10:13:39.37 &  -1:30:11.6 &  20.50 &  0.02 &   0.54 &  0.03 &   0.50 &  0.03 &   \\
VV16 & 10:13:51.21 &  -1:26:30.1 &  22.34 &  0.08 &   0.22 &  0.16 &  -0.40 &  0.10 &   \\
VV17 & 10:13:47.07 &  -1:36:04.2 &  23.89 &  0.34 &        &      &  -0.34 &  0.42 &   \\
VV18 & 10:14:17.33 &  -1:35:55.0 &  18.73 &  0.01 &   1.24 &  0.01 &   1.04 &  0.01 &   \\ 
VV19 & 10:14:24.65 &  -1:33:28.4 &  18.88 &  0.01 &   1.76 &  0.01 &   1.35 &  0.01 &   \\
VV20 & 10:14:23.18 &  -1:37:13.5 &  19.57 &  0.01 &   0.37 &  0.01 &   0.23 &  0.01 &   \\
VV21 & 10:14:13.48 &  -1:30:34.2 &  20.52 &  0.02 &   0.58 &  0.02 &   0.40 &  0.03 &   \\
VV22 & 10:14:25.84 &  -1:36:48.3 &  21.58 &  0.04 &   2.72 &  0.04 &        &      &   \\ 
VV23 & 10:11:45.89 &  -1:42:07.8 &  20.17 &  0.01 &   0.31 &  0.02 &   0.30 &  0.02 &   \\
VV24 & 10:12:10.40 &  -1:44:54.4 &  19.20 &  0.01 &   0.75 &  0.01 &   0.62 &  0.01 &   \\ 
VV25 & 10:12:30.15 &  -1:40:48.7 &  19.92 &  0.01 &   0.79 &  0.02 &   0.45 &  0.02 &   \\
VV26 & 10:12:14.65 &  -1:49:20.2 &  20.18 &  0.01 &   0.46 &  0.02 &   0.31 &  0.02 &   \\
VV27 & 10:12:50.71 &  -1:38:33.0 &  19.50 &  0.01 &   0.59 &  0.01 &   0.48 &  0.01 &    6     ACF \\
VV28 & 10:13:00.14 &  -1:42:05.4 &  20.17 &  0.01 &   0.44 &  0.02 &   0.38 &  0.02 &   36    RRab \\
VV29 & 10:12:51.48 &  -1:39:36.6 &  20.38 &  0.02 &   0.52 &  0.03 &   0.43 &  0.03 &   41    RRab \\
VV30 & 10:12:59.59 &  -1:39:00.6 &  22.76 &  0.12 &        &      &   0.47 &  0.18 &   \\
VV31 & 10:13:28.59 &  -1:45:10.2 &  19.49 &  0.01 &   1.03 &  0.01 &   0.80 &  0.02 &   \\
VV32 & 10:13:24.62 &  -1:44:54.3 &  19.83 &  0.01 &   0.63 &  0.02 &   0.36 &  0.02 &   \\
VV33 & 10:13:30.11 &  -1:41:00.1 &  20.49 &  0.02 &   0.49 &  0.03 &   0.27 &  0.03 &   28     RRc \\
VV34 & 10:13:11.14 &  -1:37:17.2 &  20.52 &  0.02 &   0.55 &  0.03 &   0.42 &  0.03 &   13    RRab \\
VV35 & 10:13:39.24 &  -1:43:10.3 &  20.20 &  0.01 &   0.36 &  0.02 &   0.23 &  0.02 &   22     RRc \\
VV36 & 10:14:02.65 &  -1:37:59.0 &  18.39 &  0.01 &   1.00 &  0.01 &   0.97 &  0.01 &   \\
VV37 & 10:14:20.48 &  -1:42:35.5 &  20.23 &  0.01 &   0.36 &  0.03 &   0.45 &  0.02 &   \\
MV01 & 10:13:10.86 &  -1:39:59.6 &  20.54 &  0.02 &   0.44 &  0.03 &   0.31 &  0.03 &   12     RRc \\
MV02 & 10:12:54.71 &  -1:31:48.6 &  19.33 &  0.01 &   0.62 &  0.01 &   0.50 &  0.01 &   19     ACH \\
MV03 & 10:12:57.98 &  -1:36:24.1 &  20.77 &  0.02 &   0.62 &  0.04 &   0.53 &  0.04 &   20    RRab \\
MV04 & 10:13:41.47 &  -1:42:12.0 &  20.23 &  0.01 &   0.47 &  0.02 &   0.41 &  0.02 &   24    RRab \\
MV05 & 10:13:19.82 &  -1:43:47.9 &  15.73 &  0.01 &        &      &   0.68 &  0.01 &   31  binary \\
MV06 & 10:13:22.84 &  -1:44:38.9 &  20.23 &  0.01 &   0.31 &  0.03 &   0.20 &  0.02 &   33     RRc \\
MV07 & 10:13:03.18 &  -1:40:39.2 &  20.49 &  0.02 &   0.62 &  0.03 &   0.49 &  0.03 &   37    RRab \\
MV08 & 10:13:27.64 &  -1:36:55.9 &  20.08 &  0.01 &   0.45 &  0.02 &   0.36 &  0.02 &   38    RRab \\
MV09 & 10:14:00.78 &  -1:40:45.9 &  19.65 &  0.01 &   0.15 &  0.02 &   0.19 &  0.01 &   43    RRab \\
MV10 & 10:13:08.42 &  -1:46:48.7 &  20.73 &  0.02 &   0.53 &  0.03 &   0.42 &  0.03 &   45    RRab \\
MV11 & 10:13:01.47 &  -1:47:30.4 &  17.82 &  0.01 &   2.21 &  0.01 &   1.60 &  0.01 &   46 reddest \\
MV12 & 10:13:43.00 &  -1:30:57.1 &  20.43 &  0.02 &   0.35 &  0.03 &   0.23 &  0.03 &    2     RRc \\
\enddata
\tablenotetext{a}{$BVI$ photometry is based on the short-exposure data of Feb 17, 2001.
IDs: VI for detected in both $V$ and $I$, VV for detected in $V$ only, MV for Mateo et al. (1995)'s
variable stars  not detected as variable stars in this search.
Remarks lists the variable IDs and types given by Mateo et al. (1995).}
\end{deluxetable}


\begin{deluxetable}{cccc}
\tablecaption{($I$,$(V-I)$) Fiducial sequences of the Sextans Dwarf Spheroidal Galaxy.\label{tbl-7}}
\tablewidth{0pc}
\tablehead { \colhead{$I$} & \colhead{$(V-I)$} & \colhead{$I$} & \colhead{$(V-I)$} }
\startdata
(RHB) & &      19.40 & 0.92 \\
19.51 & 0.80 & 19.75 & 0.89\\
19.68 & 0.65 & 19.90 & 0.88 \\
(BHB) & &      20.10 & 0.88 \\
20.09 & 0.27 & 20.30 & 0.87\\
20.20 & 0.18 & 20.50 & 0.85 \\
20.32 & 0.12 & 20.70 & 0.85 \\
20.53 & 0.07 & 20.90 & 0.84 \\
20.80 & 0.06 & 21.10 & 0.83 \\
(RGB+MS) & &   21.30 & 0.83 \\
15.88 & 1.38 & 21.50 & 0.80 \\
16.25 & 1.30 & 21.70 & 0.80\\
16.64 & 1.23 & 21.90 & 0.77 \\
16.96 & 1.17 & 22.05 & 0.75 \\
17.29 & 1.13 & 22.35 & 0.66 \\
17.61 & 1.08 & 22.65 & 0.59 \\
17.97 & 1.04 & 22.95 & 0.57 \\
18.34 & 1.01 & 23.25 & 0.56 \\
18.75 & 0.97 & 23.55 & 0.58 \\
19.09 & 0.95 & 23.85 & 0.60 \\
%
\enddata
\end{deluxetable}




\begin{deluxetable}{lccl}
\tablecaption{ A List of Metallcity Estimates for the Sextans Dwarf Spheroidal Galaxy \label{tbl-8}}
\tablewidth{0pc}
\tablehead {\colhead{Author} & \colhead{Mean [Fe/H]}  & \colhead{$\sigma$[Fe/H]}  &
\colhead{Remark}}
\startdata
Mateo et al. (1991)    & $-1.6\pm0.2$    &       & $BV$ photometry  \\
Da Costa et al. (1991) & $-1.7\pm0.25$   &       & low resolution spectroscopy \\
                       &                 &       & N=6 \\
Suntzeff et al. (1993)  & $-2.05\pm0.04$  & $0.19\pm0.02$  & low resolution spectroscopy  \\
                       &              &         &  N=43, $-2.5<$[FeH]$<-1.5$\\
Mateo et al. (1995)    & $-1.6$          &       & $BV$ photometry  \\
Geisler \& Sarajedini (1996) & $-2.0\pm0.1$ &0.17& Washington photometry   \\
Mateo (1998)              & $-1.7\pm0.2$ & $0.2\pm0.05$ & summary   \\
Shetrone et al. (2001)    & $-2.07\pm 0.10$ & 0.21  &  high resolution spectroscopy \\
                       &                 &       &  N=5, $-2.85<$[FeH]$<-1.45$ \\
Bellazzini et al. (2001) & --1.8 and --2.5 &    & $B$ photometry of the RGB bumps  \\
Rizzi et al. (2001) & $-1.91\pm0.16$ & 0.2 & $BVI$ photometry \\
This study                & $-2.1\pm0.1$ & 0.2 & $VI$ photometry  \\
\enddata
\end{deluxetable}

\begin{deluxetable}{ccccccc} 
\tablecaption{ Radial Variation of Stellar Populations in the Sextans Dwarf Spheroidal Galaxy. \label{tbl-9}}
\tablewidth{0pc}
\tablehead {
\colhead{R[arcmin]} & \colhead{N(Blue RGB)}& \colhead{N(Red RGB)}& \colhead{N(Blue SGB)}& \colhead{N(Red SGB)} & \colhead{N(MS)} }
\startdata
0--7.3   & 47 & 53 & 100 & 141 & 216 \\
7.3--11  & 54 & 39 & 142 & 136 & 223 \\
11--15   & 76 & 26 & 152 & 106 & 221 \\
15--22.5 & 65 & 37 & 168 &  92 & 215 \\
\hline
R[arcmin] & N(BHB) & N(RR) & N(RHB) & N(Bright BS) & N(Faint BS) \\
\hline
0--7.3   &  9 & 19 & 37 & 27 & 36 \\
7.3--11  &  4 & 18 & 29 & 19 & 37 \\
11--15   & 14 & 23 & 27 & 12 & 47 \\
15--22.5 &  10 & 23 & 21 &  9 & 44 \\
\hline
R[arcmin] & N(AGB bump) & N(RGB bump) & N(MWG) & HB Index \\
\hline
0--7.3   &  7 & 20 &  51 & -0.43 & \\
7.3--11  & 14 &  9 &  54 & -0.49 & \\
11--15   &  8 & 15 & 100 & -0.20 & \\
15--22.5 &  4 & 17 & 177 & -0.20 & \\
\enddata
\end{deluxetable}

\begin{deluxetable}{cc}
\tablecaption{ K-S Test of Stellar Populations in the Sextans Dwarf Spheroidal Galaxy. \label{tbl-10}}
\tablewidth{0pc}
\tablehead {\colhead{Populations} & \colhead{K-S Probability}}
\startdata
BHB -- RHB & 5.7 \% \\
Bright BS -- Faint BS & 0.4 \% \\
Blue RGB -- Red RGB & $3.98\times 10^{-9}$ \% \\
RGB bump  -- AGB bump  & 21.3 \% \\
RHB -- Bright BS  & 57.0 \%\\
BHB -- Faint BS & 51.7 \%\\
BHB -- AGB bump & 89.0 \% \\
RHB -- RGB bump  & 75.1 \% \\
\enddata
\end{deluxetable}
\clearpage



\begin{figure}
\caption{ 
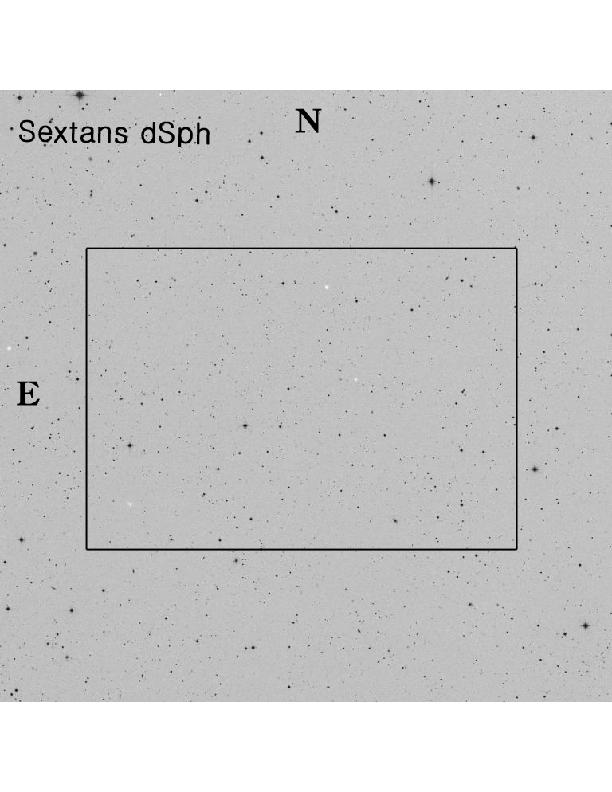: A finding chart of $1^\circ \times 1^\circ$ for the Sextans dSph in the digitized Palomar Sky Survey map.
The box  represents our observed CFH12K field of $42' \times 28'$.
North is up and east is to the left.
\label{fig1}}
\end{figure}

\begin{figure}
\plotone{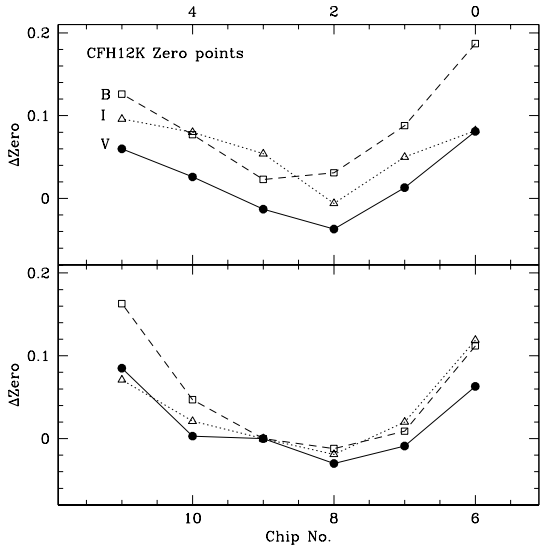}
\caption{Standard calibration zero points for the chips in
the CFH12K camera for $B$, $V_{vi}$ and $I$ filters.
The zero points are normalized with respect to that of the Chip 9.
The array of chip numbers represents the real position of the chips in the camera.
\label{fig2}}
\end{figure}

\begin{figure}
\plotone{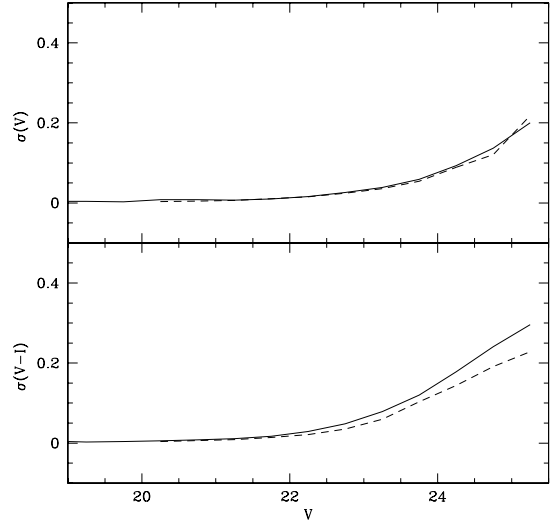}
\caption{Mean errors of our photometry. The solid line represents the
internal DAOPHOT errors, and the dashed line represents the mean differences
between the input magnitude and the output magnitude of the artificial stars.
\label{fig3}}
\end{figure}

\begin{figure}
\plotone{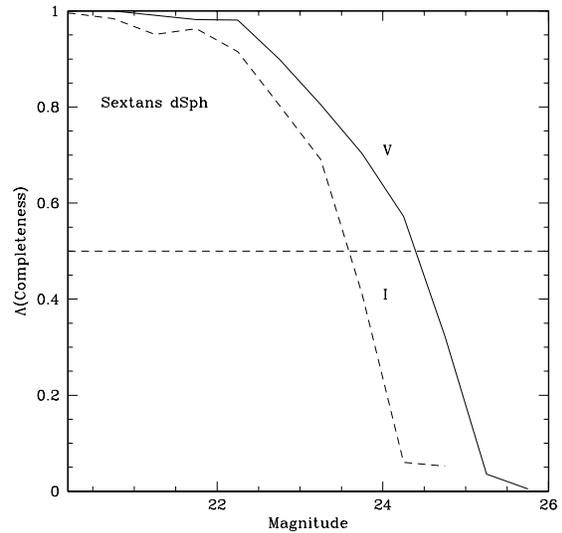}
\caption{Completeness of our $V$ (solid line) and $I$(dashed line) photometry.
The horizontal dashed line represents
the 50\% level. \label{fig4}}
\end{figure}

\begin{figure}
\plotone{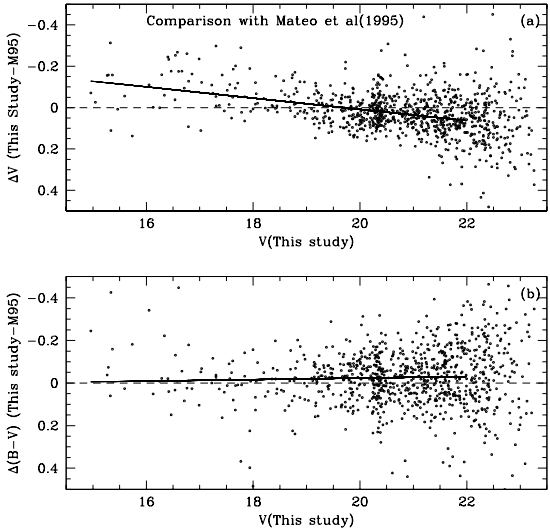}
\caption{Comparison of our $BV$ photometry with that of Mateo et al.(1995) who covered
$18'\times 18'$ field.
The solid lines represent, respectively, the linear fits of $\Delta V$ (a) and $\Delta(B-V)$
(b) for the magnitude range of $V<22$ mag.
\label{fig5}}
\end{figure}

\begin{figure}
\epsscale{0.8}
\caption{Lee.fig6.gif: $V$--($V-I$) diagrams of about 23,800 measured stars in the Sextans dSph (a) and
in a control field ($33'$ south of Pal 3) which is located at $\approx3$ degrees from the Sextans dSph
(Sohn et al. 2003) (b).
 \label{fig6}}
\end{figure}

\begin{figure}
\caption{Lee.fig7.gif: Differences in $V$ magnitude between short (Feb 17, 2001) and
long (Feb 16, 2001) exposures.
The squares represent variable star candidates.
A large number of the variable star candidates with $19.5<V<21$ mag are mostly
RR Lyraes. \label{fig7}}
\end{figure}

\begin{figure}
\caption{Lee.fig8.gif: $V$--($V-I$) diagram (a) and $V$--($B-V$)  diagram (b)
showing the variable star candidates (squares) and
the known variable stars given by Mateo et al. (1995)
(red triangles for RR Lyraes and blue triangles for anomalous Cepheids).
Two large boxes labeled with RR and AC in (a) represent roughly the regions for RR Lyraes and anomalous Cepheids,
respectively. \label{fig8}}
\end{figure}

\begin{figure}
\plotone{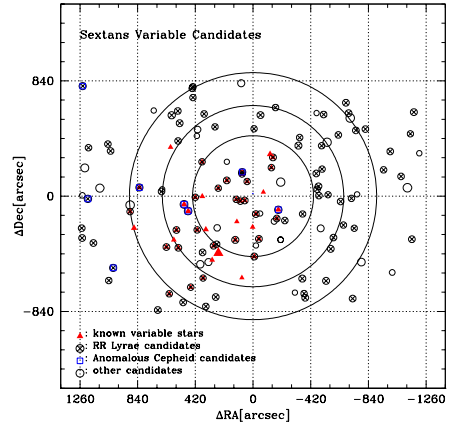}
\caption{Finding chart for the variable star candidates (small circles) and
the known variable stars given by Mateo et al.(1995)
(triangles). The size of the symbols is inversely proportional to the $V$ magnitude of the stars.
The big circles represent the boundaries at $7.3$, $11$, and $15$ arcmin.
The small open circles with cross represent the RR Lyrae candidates, the squares represent
the anomalous Cepheid candidates, and the open circles represent
other kinds of variable star candidates.\label{fig9}}
\end{figure}

\begin{figure}
\caption{Lee.fig10.gif: Fiducial sequences of the Sextans dSph in the $V$--($V-I$) diagram.
The contours represent the number density for stars with $V>23.0$ mag
which are plotted by dots.
 The error bars in the leftside represent the mean photometric errors.
Two dashed lines covering the fiducial line represent the boundary of
 $\Delta(V-I)=\pm0.1$ mag. Variable star candidates are not plotted here.
 \label{fig10}}
\end{figure}

\begin{figure}
\caption{Lee.fig11.gif: $I$--($V-I$) diagram of the measured stars in the Sextans dSph. Variable star
candidates are not plotted here. The red solid line represents the fiducial sequence of
the Sextans dSph, and the other lines represent the fiducial sequences of the
Galactic globular clusters (blue solid line: M92, left black dashed line: M15,
right black dashed line: NGC 6397,  and thin black solid line: M3).
\label{fig11}}
\end{figure}

\begin{figure}
\plotone{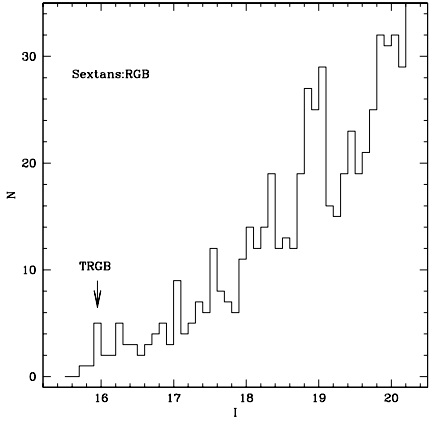}
\caption{$I$-band luminosity function of the red giants in the Sextans dSph.
The arrow represents the tip of the RGB (TRGB)
where the number of stars increases suddenly as the magnitude increases. \label{fig12}}
\end{figure}

\begin{figure}
\plotone{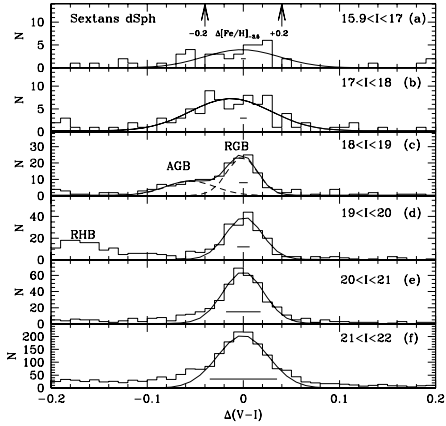}
\caption{$(V-I)$ color distribution of the red giant stars in the Sextans dSph according
to the magnitude range.
 $\Delta(V-I)$ represents the color difference with respect to the fiducial sequence.
 The horizontal bars at $\Delta(V-I)=0$ represent the mean $(V-I)$ errors.
 Two arrows in (a) represent the metallicity differences of $\Delta$[Fe/H]$=-0.2$ and +0.2 dex,
 with respect to $\Delta(V-I)=0$.
 \label{fig13}}
\end{figure}

\begin{figure}
\plotone{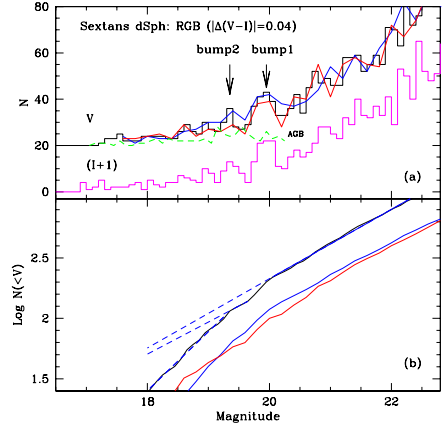}
\caption{(a) $V$-band and $I$-band differential luminosity functions of the red giant branch in the Sextans dSph
in the linear scale. 
The black histogroam represents all RGB stars for $V$-band,
 and the magenta histogram for $I$-band).
The X-axis represents $V$ and $(I+1)$ magnitude.
$I$-band luminosity functions were vertically shifted arbitrarily.
The blue, red and green solid lines represent, respectively, $V$-band luminosity functions of
the blue RGB, red RGB, and AGB.
Two arrows represent bump 1 and bump 2.
(b) Cumulative $V$-band luminosity functions.
The black, blue and red solid lines represent, respectively, $V$-band luminosity functions of
all RGB stars, the blue RGB, and the red RGB.
The dashed lines in (b) represent the linear fits to the $V$-band luminosity
function of all RGB stars, for three magnitude ranges
covered by the solid lines ($18.5<V<19.35$ mag, $19.35<V<19.95$ mag,
and $19.95<V<22.5$ mag). 
\label{fig14}}
\end{figure}

\begin{figure}
\plotone{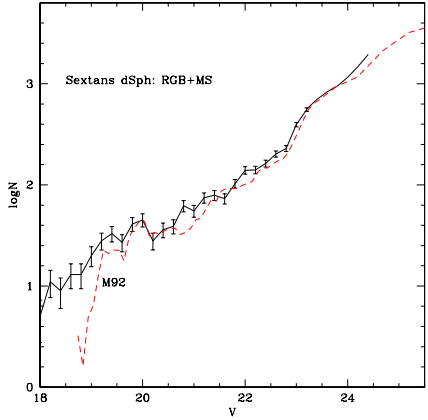}
\caption{$V$-band logarithmic luminosity function of the RGB and the MS
in the Sextans dSph (solid line)
in comparison with that of the metal-poor Galactic globular cluster M92 ($5<R<24$ arcmin)
given by Lee, K. et al. (2003) (dashed line).
The latter was arbitrarily shifted to match the former at $V=23.5$ mag.
Contributions of the field stars were subtracted and the incompleteness was
corrected.
\label{fig15}}
\end{figure}

\begin{figure}
\plotone{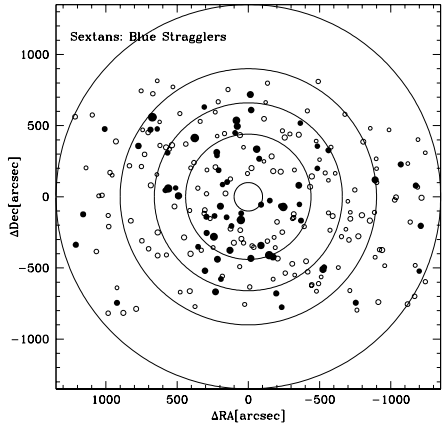}
\caption{Spatial distribution of the blue stragglers. The size of the small circles
is inversely proportional to the $V$ magnitude of the blue stragglers
(filled circles for $21.1<V<22.3$ mag, and open circles for $22.3<V<23.0$ mag).
The radii of the big circles are 100, 440, 660, 900 and 1350 arcsec.
Note that the bright blue stragglers are more strongly centrally concentrated
than the faint blue stragglers.
\label{fig16}}
\end{figure}

\begin{figure}
\plotone{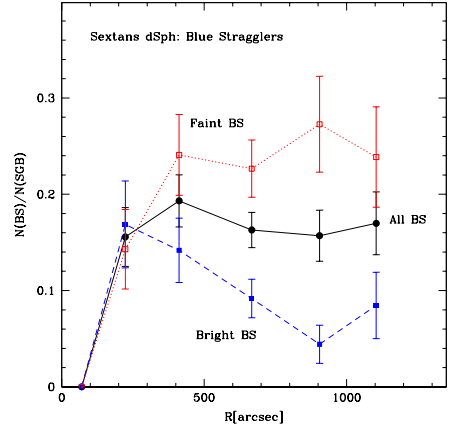}
\caption{Radial distribution of the blue stragglers.
N(BS)/N(SGB) represents the number ratio of the blue stragglers and the SGB in
the same magnitude range ($21.1<V<23.0$ mag).
Filled circles for all blue stragglers, filled squares for bright blue stragglers (with $21.1<V<22.3$ mag),
and open squares for faint blue stragglers(with $22.3<V<23.0$ mag).
Note that there is found no blue straggler at $R<100$ arcsec.
\label{fig17}}
\end{figure}

\begin{figure}
\plotone{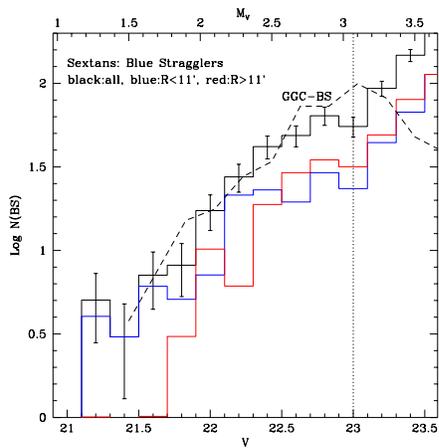}
\caption{$V$-band luminosity functions of the blue stragglers (black histogram for all,
blue histogram for blue stragglers at $R<11$ arcmin, and red historgram for blue stragglers at
$R>11$ arcmin). The black dashed line represents the combined luminosity functions (arbitrarily scaled)
of the blue stragglers in the Galactic globular clusters given by Fusi Pecci et al. (1992).
The dotted line at $V=23$ mag represents the lower magnitude boundary of the blue stragglers.
The faint end at $V>23$ mag is mainly due to the MS stars in the Sextans dSph.
\label{fig18}}
\end{figure}

\clearpage

\begin{figure}
\caption{Lee.fig19.gif: Theoretical isochrones overlayed in  the $V$--($V-I$) diagram of the Sextans dSph.
The curved lines with number labels represent the Padova isochrones for Z=0.0004 
and ages of 12.6, 10.0, 7.9 (a), 4.0, 2.0 and 1.0 Gyr (b) 
from the bottom to top.
The thick magenta  lines represent the fiducial sequence of the Sextans
dSph.
The thick blue line in (a) represents the fiducial sequence of M92.
Isochrones and the fiducial sequence of M92 were shifted according to
the distance and reddening of the Sextans dSph.
The open squares represent the variable star candidates.
\label{fig19}}
\end{figure}

\begin{figure}
\caption{Lee.fig20.gif: $V$--($V-I$) diagrams of the four regions at different radii from the center
of the Sextans dSph ($R<7.3$ arcmin, $7.3<R<11$ arcmin, $11<R<15$ arcmin, and $R>15$ arcmin).
The red solid lines represent the fiducial lines of the Sextans dSph and the boundary
with $\Delta(V-I)=\pm0.1$ mag. The red open squares represent the variable star candidates.
Boxes represent the regions for selected stellar populations.
\label{fig20}}
\end{figure}

\begin{figure}
\plotone{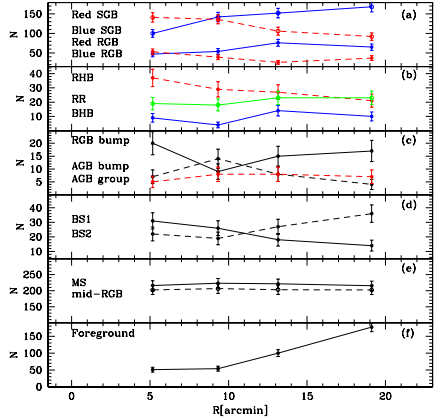}
\caption{
Radial variation of the number of stellar populations in the Sextans dSph:
(a) the blue and red SGBs, and the blue and red RGBs;
(b) the BHB (blue), RR Lyrae candidates (green), and the RGB (red);
(c) the RGB bump 1 (solid line), bump 2 (dashed line), and the AGB group (red);
(d) the bright blue stragglers (BS1, solid line) and the faint blue stragglers (BS2, dashed line);
(e) MS ($23.2<V<23.5$ mag) and mid RGB; and
(f) foreground stars ($17<V<22$ mag and $1.5<(V-I)<2.5$).
\label{fig21}}
\end{figure}

\begin{figure}
\plotone{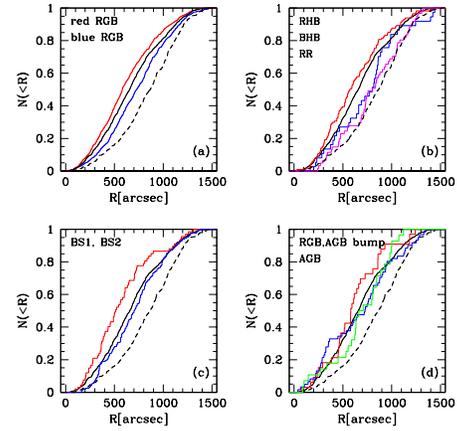}
\caption{
Cumulative number distributions of various kinds of stars
 in the Sextans dSph: (a) the red RGB (red line) and the blue RGB (blue line),
 (b) the RHB (red line), the BHB (blue line) and RR Lyrae candidates (magenta line),
 (c) the bright BSs (red line) and the faint BSs (blue line), and
 (d) the RGB bump (blue line) and the AGB bump  (red line) and the AGB group (green).
The black solid and dashed lines, respectively, represent the MS and foreground stars.
\label{fig22}}
\end{figure}

\begin{figure}
\plotone{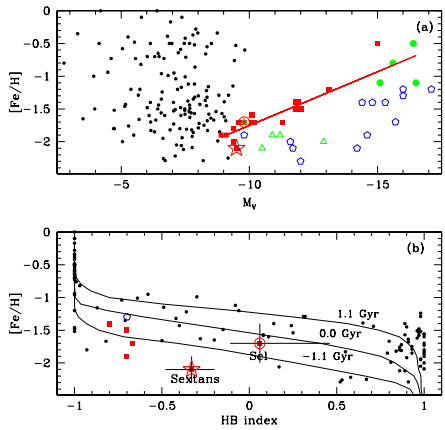}
\caption{
(a) [Fe/H] versus $M_V$ relation for dwarf galaxies in the Local Group.
(filled squares for dSphs, open triangles for transition galaxies between dSph and dwarf
irregular galaxies, open pentagons for dwarf irregular galaxies, and filled circles for
dwarf elliptical galaxies).
The small dots represent the Galactic globular clusters. The Sextans dSph and the Sculptor dSph
are marked by the starlet and open circle, respectively.
(b) [Fe/H] versus HB index relation for dwarf galaxies in the Local Group
in comparison with Galactic globular clusters. The solid lines represent the
theoretical isochrones with relative ages 1.1, 0.0, and --1.1 Gyrs (from top to bottom).
\label{fig23}}
\end{figure}

\end{document}